\documentclass[reprint,amsmath,amssymb,aps,]{revtex4-2}

\usepackage{amsmath, amssymb}
\usepackage{graphicx}
\usepackage{dcolumn}
\usepackage{bm}
\usepackage{gensymb}
\usepackage{comment}
\usepackage[normalem]{ulem}
\usepackage{hyperref}

\usepackage{color}

\begin{document}

\title{Simulating inverse patchy colloid models}

\author{Daniele Notarmuzi}
\affiliation{Institut für Theoretische Physik, TU Wien, Wiedner Hauptstraße 8-10, A-1040 Wien, Austria}
\author{Silvano Ferrari}
\affiliation{Institut für Theoretische Physik, TU Wien, Wiedner Hauptstraße 8-10, A-1040 Wien, Austria}
\author{Emanuele Locatelli}
\affiliation{Department of Physics and Astronomy, University of Padova, via F. Marzolo 8, 35131, Padova, Italy and INFN, Sezione di Padova, Via Marzolo 8, I-35131 Padova, Italy}
\author{Emanuela Bianchi}
\email{emanuela.bianchi@tuwien.ac.at}
\affiliation{Institut f\"{u}r Theoretische Physik, TU Wien, Wiedner Hauptstraße 8-10, A-1040 Wien, Austria and CNR-ISC, Uos Sapienza, Piazzale A. Moro 2, 00185 Roma, Italy}


\begin{abstract}
Nano- to micro-sized particles with differently charged surface areas exhibit complex interaction patterns, characterized by both opposite-charge attraction and like-charge repulsion. While several successful models have been proposed in the literature to describe directional attraction, models accounting for both directional attraction and directional repulsion are much less numerous and often tailored to specific microscopic systems. Here we present a simple and versatile patchy model, where the interaction energy of a pair of particles is a sum of interactions between sites of different types located within the particle volume.  We implement different formulations of this model in both a self-developed Monte Carlo code and the widely used LAMMPS Molecular Dynamics simulation software, providing basic toolkits for both simulation methods and, in the latter case, for different algorithms. By comparing physical observables and code performances, we discuss the different models, methods, and algorithms, offering insights into optimization strategies and tricks of trade.
\end{abstract}

\maketitle

\section{Introduction}

Colloids with surface regions, or "patches," characterized by different properties are commonly referred to as ``patchy colloids" and have the ability of forming specific and directional bonds thanks to the selective interactions promoted by their patches. As their synthesis at the nano- to micro-scale is nowadays achievable in a broad variety of fashions~\cite{patchyrevexp,patchyrevexp_new,colloidrevexp}, they have become viable building blocks for Materials Science applications. As such, the investigation of their large scale behavior by means of many body simulations is a crucial tool to predict and describe the broad potentialities of this class of systems~\cite{patchyrevtheo,patchyrevtheo_new}.

Models to describe colloids carrying mutually attractive patches -- here referred to as ``conventional'' patchy colloids -- have been introduced in the literature about twenty years ago~\cite{KernFrenkel2003,Glotzer2004,Bianchi2006,Doye2007} and still constitute the reference framework of many numerical investigations~\cite{Tavares2017,Chakrabarti2018,Karner-Nanolett2019,Orlandini2020,Romano2021,Sciortino2021,karner2024anisotropic,Sulc_2024} for a large variety of systems, from colloidal molecules selectively coated with ligands to spherical colloids with hydrophobic/hydrophilic patches up to functionalized all-DNA nano-structures. In contrast, the interest for particles with differently charged surface areas has been steadily growing over the last ten years -- either within the framework of rational materials design or in connection to biological systems~\cite{Bianchi2011,Podgornik2013,bianchi:2d2013,bianchi:2d2014,ismene:2014,Yura2015-2,Yigit15a,Yigit15b,Stipsitz2015,Yura2015-1,JPCM-2015-1,JPCM-2015-2,vanOostrum-2015,Cruz2016,Blanco2016,Hiero2016,Yigit2017,Araujo2017,Cruz2017,Lund2017,ipc-manner2017,ipc-cocis2017,Sabapathy-2017,Boker-2018,Boker-2018bis,Bozic2018,Bozic2018bis,silvanoepje2018,locatelli:2018,Kimura-2019,Boker-2019,Boker-2020,Videcoq-2019-exp,Videcoq-2019,Swan-2019,Jadhao-2020,Vashisth-2021,Mathews-2021,Mathews-2021bis,Shanmugathasan2022Silica,Virk2023Synthesis,popov2023jpcb,notarmuzishort,notarmuzilong}. 
Models for charged patchy colloids are intrinsically more complex than conventional patchy ones as their directional interactions must feature both attraction (between regions of opposite charge) and repulsion (between like-charged areas). As a consequence, while several established toy models for conventional patchy colloids exist and are used to describe a large variety of systems, the complexity of the charged case did not allow, so far, for the institution of a reference framework. 

Here we propose a generic model for the effective interactions between charged patchy colloids, where the particles are represented as spheres, endowed with a limited number of interaction sites, that are arranged to replicate the symmetries of a specific surface charge pattern.
The effective energy between pairs of particles is defined as the sum of contributions from each site-site interaction, for which we propose two functional forms referred to as ``overlap of spheres" (os) and ``exponential" (exp). It should be noted that the models presented here represent a generalization of the previously introduced Inverse Patchy Colloid (IPC) model~\cite{Bianchi2011} and as such we refer to them as IPC models as well. In the original formulation of the IPC model, the site-site interaction has the os functional form and its parameters are defined via a mapping to the mean-field solution of the linearized Poisson-Boltzmann equation~\cite{Bianchi2011}. The original IPC model is thus a coarse-grained representation of selected physical systems and, in turn, the parameters in the IPC model have a specific, physical meaning:
quantities such as the Debye screening length and the amount of charge carried by each interaction site must be specified in the mean-field description so to assign the site-site interaction parameters and to compute the model potential energy. 
In the present work we go beyond this specific setting, generalizing the calculation of the potential energy, 
so to include a purely parametric (toy) version of the original model, that stands as a versatile approach to the general class of colloidal systems with heterogeneous patchy interactions. By means of this generalization, these toy models are both referred to as IPC models. We implement both models in a Monte Carlo (MC) code as well as in the popular Molecular Dynamics (MD) code LAMMPS~\cite{LAMMPS}; for the latter case, we test different algorithms. We compare models, methods and algorithms looking at physical observables, as well as performances, at different thermodynamic state points and for different parameter sets.  It is worth noting that, while the IPC model can accommodate a variable number of interaction sites, we focus on particles with three sites distributed along the particle's diameter, as in Ref.~\cite{Bianchi2011}, and provide an open access toolkit to implement these systems in MC and MD-LAMMPS~\cite{IPC-toolkit}; our basic toolkit should thus be considered as an advanced starting point to simulate inverse patchy particles with possibly richer surface patterns. 

The paper is organized as follows. Detaching from the pre-existing IPC model, i.e., the coarse grained version of the mean field  potential computed analytically in Ref.~\cite{Bianchi2011}, we introduce in Section~\ref{sec:model} a general framework for parametric models based on site-site interactions. We describe their implementation in LAMMPS in Section~\ref{sec:md} and in MC in Section~\ref{sec:mc}. In Section~\ref{sec:comparison} we compare thermodynamic variables and structural properties at different state points in the fluid phase, and discuss how to optimize the performances of the different algorithms. Finally, we present our concluding remarks in Section~\ref{sec:conclusions}. 

\section{IPC general model}\label{sec:model}

IPCs are spherical particles of radius $\sigma_c$ with a fixed number of interaction sites in their interior. The off-center sites as well as the particle center -- which is the central interaction site -- are associated to different surface areas.
The arrangement of the off-center sites inside the sphere is designed to share the same symmetries of the particle surface pattern. 
Note that, while in principle the triblock pattern can feature two asymmetric patches (triblock asymmetric IPCs, ta-IPCs), in the following we focus on the symmetric case where the patches are identical in size and charge (triblock symmetric IPCs, ts-IPCs).

The interaction potential between two particles $i$ and $j$ at distance $r$ and mutual orientation $\Omega$ is given by an isotropic and suitably steep repulsion at short distances and a direction-dependent potential at intermediate distances, namely 
\begin{equation}\label{eq:U}
U=
\left\{
\begin{array}{rl}
  & U^{\rm i}(r) {\hspace{3em}\rm if\hspace{0.5em}} r < 2 \sigma_c \\
  & U^{\rm a} (r,\Omega) {\hspace{1.7em}\rm if\hspace{0.5em}} 2\sigma_{c}\leq r \leq 2\sigma_{\rm c}+r_c\\
  & 0  {\hspace{4.8em}\rm if\hspace{0.5em}} r > 2\sigma_{c}+r_c
\end{array}
\right. 
\end{equation}
where $r_c$ is a suitably chosen cut-off distance which depends on the functional form used for $U^{\rm a} (r,\Omega)$.
The isotropic repulsion, $U^{\rm i}(r)$, is a hard-core potential in MC simulations, while in MD simulations it is given by~\cite{silvanoepje2018} 
\begin{equation}\label{eq:Ur}
  U^{\rm i}(r) = A \left [ \left(\frac{2\sigma_{c}}{r}\right)^{2k}-2\left(\frac{2\sigma_{c}}{r}\right)^{k}+1 \right ]
  \end{equation}
with $k=15$ and $A=500$ (in energy units). The direction-dependent potential, $ U^{\rm a} (r,\Omega)$, is defined as
\begin{equation}\label{eq:Uromega}
U^{\rm a} (r,\Omega)  = \sum_{\alpha\beta} \epsilon_{\alpha\beta}w_{\alpha\beta}(r,\Omega)
\end{equation}
where $\alpha$ and $\beta$ specify either the center or the off-center interaction sites of the $i$ and $j$ IPC, respectively; in Eq.~(\ref{eq:Uromega}), $\epsilon_{\alpha\beta}$ is the characteristic energy strength of the $\alpha\beta$ interaction type, while $w_{\alpha\beta}$ is the associated geometric weight factor. It is worth stressing that, as we are dealing here with the toy formulation of the IPC model, the $\epsilon_{\alpha\beta}$ are always constant values assigned \textit{a priori} to characterize the $\alpha\beta$ interaction type. The distance and orientation dependence of the $\alpha\beta$ interaction type is encoded in $w_{\alpha\beta}$; however, in practice, the geometric weights are analytical functions of the inter-site distance. We enforce the dependence on the relative orientation of the two IPCs by keeping the internal geometry of the interaction sites, within each IPC, fixed.

Once the set of $n$ characteristic energies ${\bm \epsilon}=\{\epsilon_{\alpha\beta}\}$ is assigned, the energy of a pair configuration AB can be calculated via Eq.~\eqref{eq:Uromega} for any given functional forms of the $w_{\alpha\beta}$. In the ts-IPC case, the $\alpha\beta$ interaction types are $(\alpha,\beta)=$ (c, c), (c, oc) or (oc, oc) for the interactions between, respectively, the centers, the center and the off-center sites and pairs of off-center sites, meaning that the energy of a pair configuration, $u^{\rm AB}$, can be explicitly written as 
\begin{equation}\label{eq:etsipc}
u^{\rm AB}  = \epsilon_{c,c}w^{\rm AB}_{c,c}  +  \epsilon_{c,oc}w^{\rm AB}_{c,oc} +  
\epsilon_{oc,oc}w^{\rm AB}_{oc,oc}.
\end{equation}
As stated above, the $w_{\alpha \beta}^{\rm AB}$ include all the $\alpha\beta$ contributions for the given AB configuration, meaning that (c, c)-type has one term, (c, oc)-type four and (oc, oc)-type four. 

\subsection{Energy values $\epsilon_{\alpha\beta}$}
The assignment of the  $\epsilon_{\alpha\beta}$ is done by selecting reference pair configurations AB where the $\alpha\beta$ interaction type is the most relevant. The specific configurations depend on the topology of the interaction sites and their number must be equal the number $n$ of distinct interaction types. It is worth stressing that the number of interaction sites does not correspond to the number of distinct interaction types. For example, ts/ta-IPCs have both three interaction sites but there are three distinct interaction types in the former case and six in the latter. Conversely, an IPC with $n$ identical patches has $n+1$ interaction sites but the interaction energy may still be computed using three distinct interaction types.

Once a set of reference configurations is selected, the set of $n$ characteristic energies ${\bm \epsilon}=\{\epsilon_{\alpha\beta}\}$ can be determined by solving the following system of equations
\begin{equation}
{\bm u} = W {\bm \epsilon}
 \label{eq:epsilons&us}
\end{equation}
where ${\bm u}=\{u^{\rm AB}\}$ is the set of pair interaction energies in $n$ different reference configurations AB and  $W=\{\omega_{\alpha\beta}^{\rm AB}\}$ is the $n \times n$ matrix of the $\alpha\beta$ geometric factors in the AB reference configurations. Note that for ts-IPCs each equation of system~\eqref{eq:epsilons&us} has the form of Eq.~\eqref{eq:etsipc}. 

\begin{figure}[t]
\begin{center}
\includegraphics[width=0.8\columnwidth]{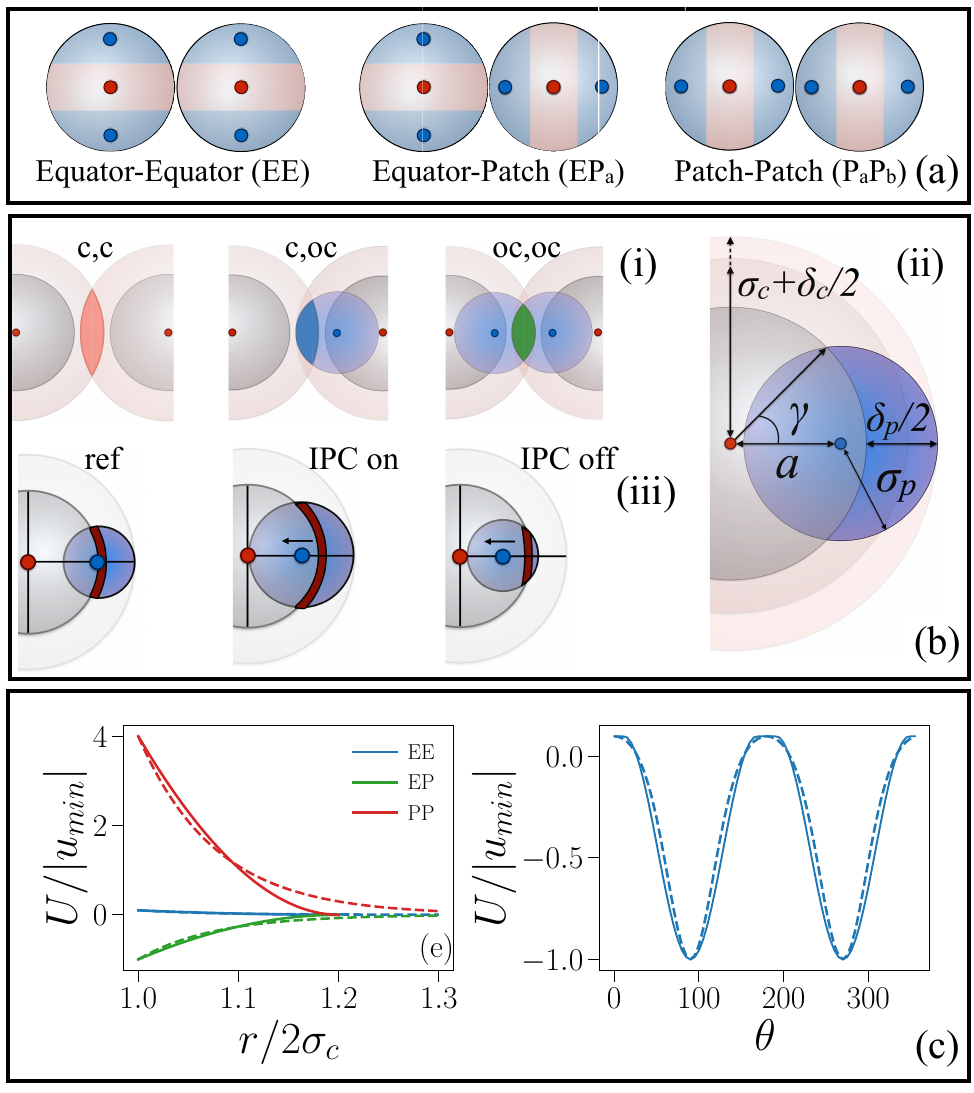}
\caption{Panel (a). Representation of possible reference configurations to estimate the characteristic interaction energies, $\epsilon_{\alpha\beta}$, between the differently charged regions on the particle surface for symmetric/asymmetric triblock inverse patchy colloids (ts/ts-IPCs). Panel (b). Sketches of the overlapping of spheres (os) model: (i) examples of overlap volumes for the center-center, center-site and site-site interaction, as labeled; (ii) geometric parameters of the os model; (iii) given a reference geometry as the one reported on the left (where the off-center site is positioned on the particle surface for simplicity), a reduction of the parameter $a$ implies an increase of the parameter $\gamma$ if the IPC-constraint is on (center), while the same reduction implies a decrease of the patch size, if the IPC-constraint is off, all other parameters being fixed (right). Panel (c) Representation of the effective potential between ts-IPCs with for the os (solid) and exp (dashed) model (parameter in Section~\ref{sec:model}C). On the left, the interaction energy $U/|u_{min}|$ is plotted versus the inter-particle distance $r/2\sigma_c$ for pairs of particles in the three mutual orientations depicted in panel (a); on the right, the interaction energy $U/|u_{min}|$ is reported for pairs of particles at contact where one particle rotates with respect to the other: at $\theta=0^{\rm o}$ the two particles are in the EE configuration, while at $\theta=90^{\rm o}$ particles are in the EP configuration.
}
\label{fig:fig1} 
\end{center}      
\end{figure}

The system of equations~(\ref{eq:epsilons&us}) requires setting the $u^{\rm AB}$-values: one can either obtain them from a mean-field description of a physical system (e.g., as in Ref.~\cite{Bianchi2011}) or by fixing them arbitrarily -- the latter choice being referred here to as toy. As we set them arbitrarily, we obtain the corresponding ${\bm \epsilon}$ by selecting AB configurations where particles are positioned at contact, i.e., at the minimum possible distance between two particles $r=2\sigma_c$, with different mutual orientations AB. For ts/ta-IPCs, on which we focus in this work, sets of possible reference configurations are reported in Figure~\ref{fig:fig1}a. The best reference configurations are the equator-equator, EE, the equator-patch, E${\rm P_{a,b}}$, and the patch-patch, ${\rm P_aP_b}$, orientations, where the subscripts $a$ and $b$ refer to possibly different patches (see Supporting Information, SI, Section 1).

\subsection{Geometric weights $w_{\alpha\beta}$}
We propose two functional forms for the geometric weights $w_{\alpha\beta}$. In the overlap of spheres (os) approach, each interaction site is associated to an interaction sphere. The geometric weight $w_{\alpha\beta}$ is then proportional to the total overlap volume between pairs of $\alpha\beta$ interaction spheres (see Figure~\ref{fig:fig1}b,i).  The analytic form of the $w_{\alpha\beta}$  are reported in Ref.~\cite{Bianchi2011} and in Section~1 of the SI for completeness. 
In the exponential (exp) approach, the geometric weights $w_{\alpha\beta}$ decay exponentially with the site-site distance. 

\subsubsection{Overlap of Spheres (os) model}

The radius of the interaction sphere of the central site is $\sigma_c+\delta_c/2$ while for the off-center sites it is $\sigma_p$; $\delta_c$ sets the center-to-center interaction range (see Figure~\ref{fig:fig1}b,ii). As, in general, an off-center site is located inside the particle or on its surface and its position is specified by an eccentricity parameter $a\leq\sigma_c$, its interaction sphere should extends outside the particle surface (i.e. $a+\sigma_p>\sigma_c$). It is then possible to define a surface patch \textit{via} the half-opening angle $\gamma$
\begin{equation}
\cos \gamma = \frac{\sigma_c^2 +a^2 -\sigma_{p}^2}{2a\sigma_c}.
\end{equation} 
Further, we define the patch interaction range $\delta_p$ as
\begin{equation}
\frac{\delta_p}{2}=a+\sigma_{p}-\sigma_{c} 
\label{eq:deltap}
\end{equation}
(see Figure~\ref{fig:fig1}b,ii). Since $\sigma_{c}$ fixes the unit of length of the model ($\sigma=2\sigma_c=1$), the parameters to be chosen are $\delta_c$, $\sigma_{p}$ and $a$.
While $\delta_c$ can be uniquely related to the experimental inter-particle interaction range, $a$ and $\sigma_{p}$ are related to both the experimental patch size and interaction range according to the aforementioned geometric constraints.  
When referring to a mean-field model for heterogeneously charged colloids or when simply postulating a common 
screening length, all interaction sites must have the same interaction range, that is determined by the electrostatic screening of the surrounding solvent. 
As a consequence, the relation $\delta_c=\delta_p=\delta$, referred to as IPC-constraint, must be imposed. In this case, the choice of $\sigma_{p}$ and $a$ defines not only $\gamma$ but also $\delta$  (see Figure~\ref{fig:fig1}b,ii). It is worth noting that, when satisfying the IPC-constraint, a change in $a$ must be accompanied by a change in $\sigma_p$, so that the patch interaction range remains equal to the particle interaction range (see Figure~\ref{fig:fig1}b,iii). In contrast, without the IPC-constraint, a change in $a$ does not imply any change in $\sigma_p$ but rather in $\delta_p$, given by Eq.~\eqref{eq:deltap}.
Notice also that the patch size $\gamma$ is affected in an opposite way by a change of $a$ (see again Figure~\ref{fig:fig1}b,iii) with respect to whether the IPC-constraint is on or off. Indeed, in the former case $\gamma$ increases upon decreasing $a$, as the constraint on the interaction range makes the whole patch increase in size. In the latter case, decreasing $a$ burrows the patch inside the colloid; as such $\gamma$ decreases.

\subsubsection{Exponential (exp) model}

In this model, we endow each interaction site with an exponentially decaying function of the site-site distance, thus
\begin{equation}
   \omega_{\alpha \beta}^{\mathrm{AB}}= \sum_{r_{\alpha\beta} | \mathrm{AB}}e^{-\kappa (r_{\alpha \beta}-r^{c}_{\alpha \beta})}
\end{equation}
where $k$ is a characteristic inverse length and $r_{\alpha \beta}^c$ the cut-off distances associated to the different $\alpha \beta$ interactions. 
The cut-off distances are defined as $r_{c, c}^c=2\sigma_c$, $r_{c, oc}^c=2\sigma_c-a$ and $r_{oc, oc}^c=2\sigma_c-2a$, while the common screening factor $\kappa$ is a free parameter of the model. Physically, it
is related to the screening length of the solution as it represents the characteristic length scale of the
interaction between charged sites. In the present work, however, we set $\kappa$ so to get the best match between the exp and os model potentials. We do so to characterize their computational efficiency and to understand whether or not these two choices of the function $w_{\alpha\beta}$ result in particle models with significantly different behavior.

\subsection{Pair potential representation} 

In Figure~\ref{fig:fig1}c we report the radial and the angular dependence of the interaction energy between sample pairs of IPCs. For both models we consider ts-IPCs with $2\sigma_c=1$, $a=0.22$ and ${\vec u}=\{u_{\rm EE}, u_{\rm EP}, u_{\rm PP}\} =\{0.1,-1.0,4.0\}$. Notice that $2\sigma_c$ and $u_{\rm EP}$ are our length and energy units. For the os model, we set  $\sigma_p=0.38$, which translates in an interaction range $\delta=0.2$, while for the exp model we set $\kappa=13$. The radial dependence is reported for each reference configuration in Fig.~\ref{fig:fig1}c,left. The angular dependence in Fig.~\ref{fig:fig1}c,right has, as starting orientation, the EE configuration and is obtained rotating one of the two particles around the axis perpendicular to the plane and passing through the center of the particle. The parameters reported in this section will be used throughout the rest of the paper. 

\section{Monte Carlo simulations}\label{sec:mc}

Monte Carlo simulations of the IPC model are performed readapting the 
publicly available code by Rovigatti et al~\cite{Rovigatti2018How} and we provide an open access toolkit to readily implement all these systems in MC~\cite{IPC-toolkit}. 
We establish our Monte Carlo simulation code on the Virtual Move Monte Carlo algorithm  (VMMC)~\cite{Whitelam2007Avoiding}, of which we give here a brief summary; for a detailed description, see
Ref.s~\cite{Rovigatti2018How, Whitelam2007Avoiding, Whitelam2009The, Ruzicka2014Collective}. 
Specifically, we consider an ``ad litteram'' implementation of the algorithm explained in
Ref.~\cite{Rovigatti2018How}. VMMC is a cluster move that works efficiently with strongly interacting particles. The algorithm builds clusters of particles dynamically, by proposing to move a randomly chosen particle, the ``seed'' of the move, 
and checking whether or not moving it would increase the energy of its neighbours. If so, said neighbours may be recruited (clustered) in the move.
Practically, a move (rotation or translation) is selected, together with the
seed of the move. The move of the seed can be a rotation or a translation, each with probability $1/2$. 
Both moves are regulated by a parameter each, the maximum angle of rotation $\phi_{max}$
and the maximum translation $\delta_{max}$.
For each one of the seed's neighbours, the pair energy is then computed 
before and after the move. Depending on the Metropolis acceptance rate specified
in Ref.~\cite{Rovigatti2018How}, the neighbor particle 
may be recruited in 
the cluster or not. If so,  
the same procedure described above is applied to the newly recruited particle,
building the cluster 
iteratively. Once there are no more particles to be recruited, the movement of the cluster as a whole
rigid body is accepted or rejected depending again on a Metropolis acceptance rate. 

We stress that the move
can be rejected by two early rejection mechanisms:
(i) if one particle of the cluster would move by a distance that is larger than $\Delta^c$ (which can only happen
in case of a cluster rotation) and (ii) 
if the number of particles recruited
in the cluster is larger than $S^c$
~\cite{Whitelam2007Avoiding}. 
The four parameters $\Delta^c$, $S^c$, $\phi_{max}$ and $\delta_{max}$ 
regulate the acceptance rate
of the algorithm, which is expected to vary significantly between the highly diluted and the 
dense phases.

Note that excessively large clusters are prevented for two reasons. First, if the recruitment procedure is left unchecked, a cluster that - under periodic
boundary conditions - contains multiple copies of the same particle  may appear and should be discarded, because it is unphysical.
Second, system-spanning clusters should be prevented, because their sole result would be a very costly rigid rotation or translation with no internal conformation rearrangement.

To this aim, we set $\Delta^c=1.8$ and $S^c=25$. 
Furthermore, we set $\delta_{max}=0.05$ and $\phi_{max}=0.1$.
These values of $\delta_{max}$ and $\phi_{max}$ correspond to 
having an average acceptance rate $A_r \simeq 0.3$ in simulations at low densities if
only single-particle roto-translations are used.

Concerning the exp model, note that the pair energy in MC simulations is cut at 0 for all distances that are sufficiently large for the interaction potential of all the references configurations to be at least $10^{-3}$ times the value at contact.

We choose the VMMC move for the present investigation as it is particularly
suited to study particles with limited bonding valence,
especially at low temperature~\cite{Whitelam2009The}, given its ability to escape from
kinetic traps that are common when the temperature is sufficiently small. The study of 
these regions of the phase diagram, in fact, is of particular interest when dealing with
patchy particles as they may show peculiar assembly and thermodynamic properties under these conditions, as it is indeed the focus of several studies~\cite{Rovigatti2018How}.
We thus implement and test a move that we believe to be useful for simulations
of IPC systems under conditions that may be hard to simulate efficiently using standard roto-translations of individual particles~\cite{Rovigatti2018How}. 

\section{Molecular Dynamics simulations with LAMMPS}\label{sec:md}

In order to implement the model introduced in section \ref{sec:model} in a MD code, specifically in LAMMPS, we consider two different approaches: a ``constrained''-MD algorithm, to simulate the IPC as a rigid body and a ``bead-spring'' algorithm, to maintain the internal arrangement of the sites using bonding and bending potentials. We will compare the two by monitoring performances, thermodynamic variables, structure and dynamic properties, using Monte Carlo simulations as an independent reference. We carry on such a comparison to provide a guideline for the reader interested in using the model, so that pros and cons of each algorithm may be evaluated for future applications. 
Setting up a simulation of IPCs in LAMMPS entails the computation of the pair potential in a suitable format and the creation of a suitable initial configuration, where the chosen IPCs arrangement is correctly implemented. Our approach consists in tabulating the site-site potentials; in practice, one needs to generate suitably
formatted files. Further, the LAMMPS’ initial data (or “data file”) should also be generated. A code for such a setup, plus other scripts useful for post-processing, is available at~\cite{IPC-LAMMPS}. In addition, an open access toolkit is available to quickly setup simulations with MD-LAMMPS of this system~\cite{IPC-toolkit}.

\subsection{Introducing IPCs' pair potentials in LAMMPS}
As mentioned, we introduce the IPCs' pair potential in LAMMPS as a set of tabulated site-site potentials.
We employ the radial dependencies of each site-site interaction: we tabulate the values of $\bar{\epsilon}_{\alpha \beta}(r) = \epsilon_{\alpha \beta}\omega_{\alpha \beta}(r)$ for a suitable range of distances. In the simulation, $r$ is taken as the distance between sites of type $\alpha$ and $\beta$, belonging to different IPCs. The use of tabulation files allows for a simple and efficient implementation in LAMMPS: we provide a code to generate said files, for any given set of values of the parameters, in a format suitable for LAMMPS's \verb|pair_style table|~\cite{{IPC-LAMMPS}}.

Notice that a cutoff distance should be provided for both the os and exp model. By construction, in the os model the potential goes to zero when the interaction spheres do not overlap anymore, i.e., at $r=2\sigma_c+\delta_c$ for $\bar{\epsilon}_{cc}$, at $r=\sigma_c+\delta_c/2+\sigma_p$ for $\bar{\epsilon}_{cp}$ and at $r=2\sigma_p$ for $\bar{\epsilon}_{pp}$. On the contrary, in the exp model we have to enforce a cutoff: we cut the pair energy to zero (i.e. we stop the tabulation of the values), when all the reference configurations provide with an interaction energy that is at least $10^{-4}$ units of energy, independently of the value at contact. 

\subsection{Rigid body vs bead-spring}

\begin{figure}[t]
\begin{center}
\includegraphics[width=0.7\columnwidth]{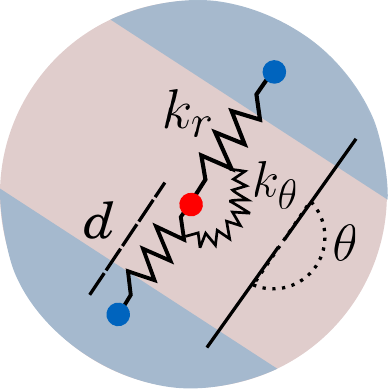}
\caption{Sketch of a bead-spring IPC, with off-center sites connected to the central site via harmonic springs of strength $k_r$. The axiality is maintained by a harmonic bending potential of strength $k_{\theta}$.  
}
\label{fig:sketch} 
\end{center}      
\end{figure}

We discuss here pros and cons of two algorithms introduced above. On the one hand, rigid bodies (``constrained-MD'') are, generally, computationally more expensive than bead-spring algorithms and require more care to be initialized properly. Further, in LAMMPS, rigid bodies are not compatible with a relatively large subset of functionalities; in addition, simulating complex arrangements with four or more off-center sites can become cumbersome. However, they allow to maintain the sites' arrangement inside the IPCs with great accuracy.\\
On the other hand, bead-spring algorithms are extremely flexible and can easily be extended to, potentially, any patch number and arrangement. They entail the definition of suitable bonding and bending potentials, that are computationally relatively inexpensive with respect to the rigid body constraints, and pertain to the sites of single IPCs, thus scale linearly with the size of the system. These potentials can be tuned to maintain, up to a certain degree, the arrangement of the IPC sites and, possibly, allow also for an easy extension to mobile sites. However, their main issue is that they are parametrical, i.e., they require to fix additional parameters; the effect of choosing a value (instead of another) may not be trivial.\\
From the algorithmic perspective, LAMMPS allows to simulate rigid bodies 
by setting up the equations of motion with Ciccotti's formulation~\cite{ciccotti}, which avoids the singularities imposed by the internal linear architecture of the particles. The resulting equations are then integrated with RATTLE~\cite{RATTLE}, an algorithm that guarantees that the coordinates and velocities of the entities within a molecule satisfy the internal geometric constraints. It is worth noting that a previous, self-developed molecular dynamic simulation code -- developed by some of the authors and described in Ref.s~\cite{silvanoepje2018,silvanoepje2022erratum} and publicly available at~\cite{IPCsim} -- uses the same two algorithms.

The bead-spring algorithm aims at being essential and parsimonious. The central site is held together with each of the off-center sites by simple harmonic springs, described by an interaction potential 
\begin{equation}
     U(d) = k_r (d-a)^2,
\label{eq:harm}
\end{equation}
where $d$ is the site-site distance and $a$ is eccentricity parameter, that is set to be the spring's rest length (see Fig.~\ref{fig:sketch}); notice that we omit the usual $1/2$ prefactor, as in the LAMMPS's implementation of this interaction. Suitable bending potentials should be employed to keep triplets of sites in the right configuration; again a minimalistic harmonic bending potential is employed  
\begin{equation}
     U(\theta) = k_{\theta} (\theta-\theta_0)^2,
\label{eq:harm_theta}
\end{equation}
where $\theta$ is the angle between a triplet of interaction sites and $\theta_0$ is the reference angle for said triplet. For a ts/ta-IPC there is only one bending angle, that is the angle between the vectors connecting the central with the two off-center sites; the reference angle is $\pi$ (see Fig.~\ref{fig:sketch}). 
As hinted previously, $k_r$ and $k_{\theta}$ are parameters to be tuned. Since we aim at simulating quasi-rigid objects, we are tempted to use very large values for both. However, as known~\cite{frenkel2023understanding}, very large spring constants cause numerical instabilities at fixed $\Delta t$; indeed, as it will be discussed in Sec.~\ref{sec:comparison}, increasing or decreasing the value of $k_{r}$ and $k_{\theta}$ does lead to consequences that are sometimes subtle. One has to choose said values carefully, fixing them one at the time looking for the optimal values that prevent distortion of internal site arrangement and avoid massive efficiency drops.  

\section{Comparison between MC and MD-LAMMPS simulation outputs}\label{sec:comparison}
We present now the comparison between MC and MD simulation results, performed at the same state points, focusing on triblock symmetric colloids (ts-IPCs): we fix the same parameters used in Fig.~\ref{fig:fig1}c and simulate $N=1000$ in a cubic box of linear size $L$ at $T=0.150$ and $\rho=0.25, \, 0.50, \, 0.75$, corresponding to $L=15.9\sigma, \, 12.6\sigma, \, 11.0\sigma$ respectively.

For MC simulations, we simulate 8.2$\cdot 10^6$ MC steps, a step being defined as the attempt to change the system's state $N$ times; for each state point, we perform 8 parallel runs. A configuration
is saved every $10^4$ MC steps; however, since the first $2 \cdot 10^6$ Monte Carlo
steps are discarded to allow for equilibration, 
we collect a total 4960 configurations per state point over which we perform our measurements. 
MC simulations start from a randomly generated configuration; at 
$\rho=0.75$ the starting configuration is obtained by melting an FCC crystal with the assigned density.

For MD simulations, we perform NVT runs, starting from an FCC crystal and melting it at temperature $T=1.000$ for $10^4$ time steps; then, we quench the system to $T=0.150$ using the same number of time steps. Finally, we simulate the system 
for $10^7$ time steps. 
We simulate both algorithms, namely the ``constrained-MD'' and the bead-spring one; we compare the effect of different thermostats, using either the Nosé–Hoover (NH) or the Langevin (LANG). While employing the NH thermostat, the dumping coefficient is always set to $T_d=100 \Delta t$ for both bead-spring and constrained-MD. For the bead-spring-NH, we considered three sets of systems, defined by the values of $k_r$, $k_{\theta}$ and of the time step $\Delta t$: (i) fixing $k_r = k_{\theta}=k$, $10^3 \leq k \leq 10^4$ 
and $\Delta t=10^{-3}$, (ii) the same values of $k$ and variable $\Delta t$, specifically,
$\Delta t=10^{-3}$ for $k=10^3$,
$\Delta t=5 \cdot 10^{-4}$ for $2\cdot 10^3 \leq k
\leq 8 \cdot 10^3$ 
and $\Delta t=10^{-4}$ for $k=10^4$ and (iii) fixing $\Delta t=10^{-3}$, $k_{r}=10^4$ and variable $100 \leq k_{\theta} \leq 10^4$. 

For bead-spring LANG simulations, we
considered $k_r=k_{\theta}=k=10^4$ and $\Delta t=10^{-3}$. For both the bead-spring and constrained-MD LANG simulations, we consider $T_d= 10.0 \tau $ and $T_d=1.0 \tau$.

First, we focus on the single particle properties, namely the axiality and the eccentricity in the harmonic bonds case; this will help us in the choice of the parameters $k_r$ and $k_{\theta}$. Once those are fixed, we look at thermodynamic quantities, such as the average temperature (and its fluctuations) and the average interaction energy (and its fluctuations), across the different algorithms. Then, we compare the structural properties of the fluid, computing the radial distribution functions and the distribution of the number of bonds per particle. Finally, we estimate the efficiency of the different simulation methods and algorithms by comparing the simulation run times of both IPC models at all the investigated state points. 

\subsection{Single-particle properties of IPCs with harmonic bods}
We investigate the effect of varying the spring constants $k_r$ and $k_{\theta}$ on the single particle properties of bead-spring IPCs. As showed in other models of patchy particles, replacing rigid with floppy bonds may lead to important differences in the phase diagram\cite{smallenburg2014erasing,smallenburg2015tuning}. Depending on the values of $k_r$ and $k_{\theta}$, significant radial as well as angular fluctuations of the off-center charges, relative to the imposed triblock topology, may happen; in the context of this work, we aim at providing the reader with a set of optimal values to simulate rigid-like IPCs that allow to maintain computational efficiency (see Sec.~5 of the SI). We look at single particle properties, specifically the distributions of the axial and angular displacements, upon varying the spring constant $k_r$ and $k_{\theta}$ in Eqs.~(\ref{eq:harm}) and (\ref{eq:harm_theta}); we further check if employing different thermostats affects the results, reporting here the NH case. We also focus here on the os model at $\rho=0.5$; data for the exp model, as well as data for different values of $\rho$ are reported in the SI. 

\begin{figure*}[t]
\begin{center}
\includegraphics[width=\textwidth]{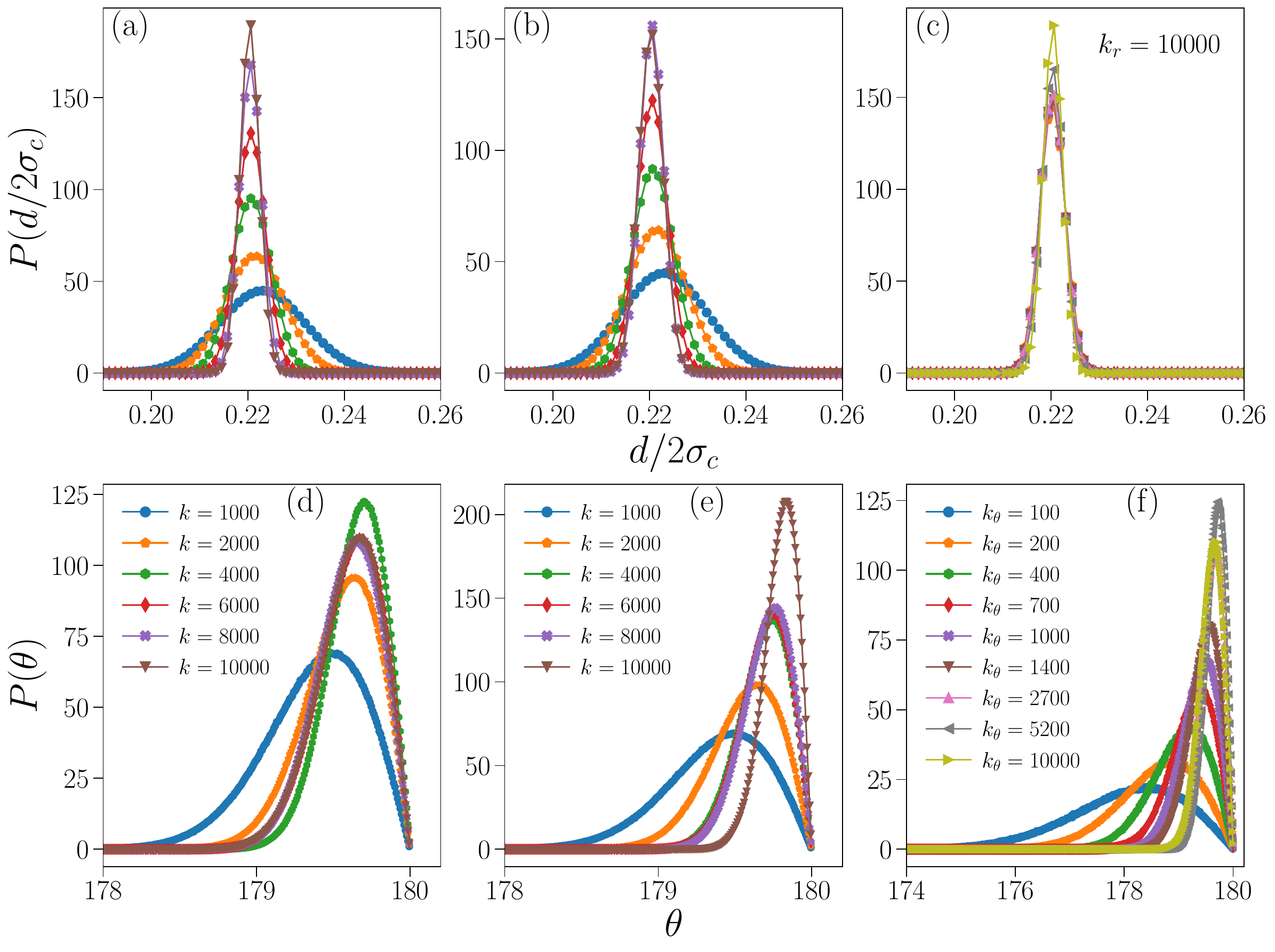}
\caption{Single particle properties for the os model at $\rho=0.50$ with the Nosé–Hoover thermostat. 
Top: eccentricity distributions.
Bottom: axial angle distributions.
(a),(d) Systems with $k_{\theta}=k_{r} \equiv k$ and $\Delta t=10^{-3}$.
(b),(e) Systems with $k_{\theta}=k_{r} \equiv k$ and variable $\Delta t$ (see Sec.~\ref{sec:comparison}) 
(c),(f) Systems with $k_{r}=10^4$, $k_{\theta}$ as specified in the legend of panel (f) and $\Delta t=10^{-3}$.
}
\label{fig:fig2} 
\end{center}      
\end{figure*}

In Fig.~\ref{fig:fig2}a-c we report the distributions of the eccentricity, i.e. of the distances between the central and the off-center sites at $\rho=0.5$; in Fig.~\ref{fig:fig2}d-f we report the distributions of the axial angle, i.e. the angle between the three sites in each IPC. We focus on the different sets (i)-(iii), described above, case (i) in Fig.~\ref{fig:fig2}a,d, case (ii) in Fig.~\ref{fig:fig2}b,e and case (iii) in Fig.~\ref{fig:fig2}c,f.

Notice that the reference methods for rigid bodies (MC or constrained-MD) would yield $\delta$-functions around the chosen value, that is, $a/\sigma=0.22$ for the distributions of the eccentricity and $\theta=\pi$ for the distributions of the axial angle. Here we omit both for simplicity. Notice also that additional data at different values of the density are reported in the SI, {Section 2}.

We start from case (i), where we fix the two spring constants $k_r$ and $k_{\theta}$ to have an equal numerical value (Fig.~\ref{fig:fig2}a,d).
We observe that a spring constant of at least $4 \cdot 10^3 u_{\rm EP}/\sigma^2$ is needed to ensure that the eccentricity is, on average, the one selected initially. Indeed, for smaller values of $k$, the average eccentricity is larger than the set value; additionally, the fluctuations are large, which can, potentially, lead to different result in the self-assembly at lower temperatures. On the other hand, we observe that, upon increasing $k$, the axiality shows a slight non-monotonic behaviour, that is also accompanied by a significant deviation from the reference mean energy for $k\leq 6 \cdot 10^3$  (see Sec.~4 of the SI). This can be resolved by decreasing the integration time step, as considered in (ii) (Fig.~\ref{fig:fig2}b,e). However, the drawback of this approach is a considerable loss of computational efficiency. A more sensible approach is case (iii) (Fig.~\ref{fig:fig2}c,f), where we decouple $k_r$ and $k_{\theta}$ and we keep $\Delta t$ fixed. We choose $k_r= 10^4 u_{\rm EP}/\sigma^2$, to minimize radial fluctuations and we vary $k_{\theta}$ between $10^2u_{\rm EP}$  and $10^4 u_{\rm EP}$ in a logarithmic fashion. Notice that the distribution of the eccentricity is minimally affected by the value of $k_{\theta}$ whereas we find a best value for $k_{\theta}$ from the distributions of the axial angle at $k_{\theta} \approx 5 \cdot 10^3 u_{\rm EP}$. As reported in the SI, we find similar results performing Langevin Dynamics simulations. We thus select $k_r= 10^4 u_{\rm EP}/\sigma^2$, $k_{\theta}=5.2 \cdot 10^3 u_{\rm EP}$ as our best candidate for bead-spring IPCs.

\subsection{Thermodynamics}

We now check the thermodynamic properties of the system, namely the kinetic temperature and the mean pair potential energy per particle, for the different model and thermostats considered.

\begin{table*}[!b]
\resizebox{\textwidth}{!}{
  \centering
  \begin{tabular}{c|cc|cc}
    \multicolumn{5}{c}{$\rho=0.25$}\\
    \multicolumn{5}{c}{}\\
    \hline
    & \multicolumn{2}{c|}{$T$} & \multicolumn{2}{c}{$U$} \\
    \hline
    & os & exp & os & exp \\
    \hline
NH, RG, $T_d=0.10$ & 0.1500 $\pm$ 0.0029 & 0.1500 $\pm$ 0.0030 & -0.6346 $\pm$ 0.0188 & -0.3118 $\pm$ 0.0121 \\
LG, RG, $T_d=1.00$ & 0.1501 $\pm$ 0.0030 & 0.1501 $\pm$ 0.0030 & -0.6339 $\pm$ 0.0179 & -0.3120 $\pm$ 0.0119 \\
LG, RG, $T_d=0.10$ & 0.1503 $\pm$ 0.0037 & 0.1502 $\pm$ 0.0035 & -0.6315 $\pm$ 0.0242 & -0.3114 $\pm$ 0.0135 \\
NH, $k_{\theta}=5.2 \cdot 10^3$, $T_d=0.10$ & 0.1500 $\pm$ 0.0023 & 0.1500 $\pm$ 0.0023 & -0.5570 $\pm$ 0.0162 & -0.2737 $\pm$ 0.0153 \\
    \hline
    \multicolumn{5}{c}{}\\
    \multicolumn{5}{c}{$\rho=0.50$}\\
    \multicolumn{5}{c}{}\\
    \hline
    & \multicolumn{2}{c|}{$T$} & \multicolumn{2}{c}{$U$} \\
    \hline
    & os & exp & os & exp \\
    \hline
NH, RG, $T_d=0.10$ & 0.1500 $\pm$ 0.0030 & 0.1501 $\pm$ 0.0030 & -0.9370 $\pm$ 0.0156 & -0.5768 $\pm$ 0.0132 \\
LG, RG, $T_d=1.00$ & 0.1502 $\pm$ 0.0030 & 0.1501 $\pm$ 0.0031 & -0.9362 $\pm$ 0.0161 & -0.5765 $\pm$ 0.0132 \\
LG, RG, $T_d=0.10$ & 0.1505 $\pm$ 0.0039 & 0.1504 $\pm$ 0.0037 & -0.9332 $\pm$ 0.0234 & -0.5755 $\pm$ 0.0167 \\
NH, $k_{\theta}=5.2 \cdot 10^3$, $T_d=0.10$ & 0.1500 $\pm$ 0.0023 & 0.1500 $\pm$ 0.0023 & -0.8600 $\pm$ 0.0148 & -0.5097 $\pm$ 0.0153 \\
    \hline
    \multicolumn{5}{c}{}\\
    \multicolumn{5}{c}{$\rho=0.75$}\\
    \multicolumn{5}{c}{}\\
    \hline
    & \multicolumn{2}{c|}{$T$} & \multicolumn{2}{c}{$U$} \\
    \hline
     & os & exp & os & exp \\
    \hline
NH, RG, $T_d=0.10$ & 0.1500 $\pm$ 0.0030 & 0.1500 $\pm$ 0.0030 & -1.2031 $\pm$ 0.0134 & -0.8423 $\pm$ 0.0127 \\
LG, RG, $T_d=1.00$ & 0.1502 $\pm$ 0.0030 & 0.1501 $\pm$ 0.0030 & -1.2023 $\pm$ 0.0139 & -0.8418 $\pm$ 0.0128 \\
LG, RG, $T_d=0.10$ & 0.1506 $\pm$ 0.0040 & 0.1505 $\pm$ 0.0038 & -1.1987 $\pm$ 0.0206 & -0.8398 $\pm$ 0.0174 \\
NH, $k_{\theta}=5.2 \cdot 10^3$, $T_d=0.10$ & 0.1500 $\pm$ 0.0022 & 0.1500 $\pm$ 0.0022 & -1.1344 $\pm$ 0.0134 & -0.7594 $\pm$ 0.0144 \\
\hline
\end{tabular}
}
  \caption{Average kinetic temperature and pair energy per particle in LAMMPS simulations.}
  \label{tab:TU}
\end{table*}

The results are reported in Tables~\ref{tab:TU}; more data are reported in the SI. The kinetic temperature is always compatible with the temperature of the heath bath, both considering NH and LANG thermostats; in the latter case, we also show that slightly changing the damping coefficient does not affect the thermodynamics, as should be the case. In general, the absolute value of the potential energy per particle increases upon increasing the density, as expected in a more dense liquid. Interestingly, the exp model is characterized by a smaller absolute value of the  potential energy, with respect to the os model, even though the two have, by construction, the same interaction energy at contact, in the reference configurations. However, the exp model, as highlighted in Fig.~\ref{fig:fig1}c, has a longer range than the os model: as such, the repulsive PP and EE contributions.
Finally, the pair potential energy per particle is, for the bead-spring parameters selected, systematically smaller than the rigid counterpart (6-12\%). As we will see in the next section, this small discrepancy is accompanied by small differences in the local structure of the fluid that, overall, remain of minor importance.

\subsection{Fluid structure and network properties}

We now look at the structure of the fluid at all length scales, focusing on its immediate neighborhood first and then considering the full radial distribution function.

\begin{figure*}[t]
\begin{center}
\includegraphics[width=\textwidth]{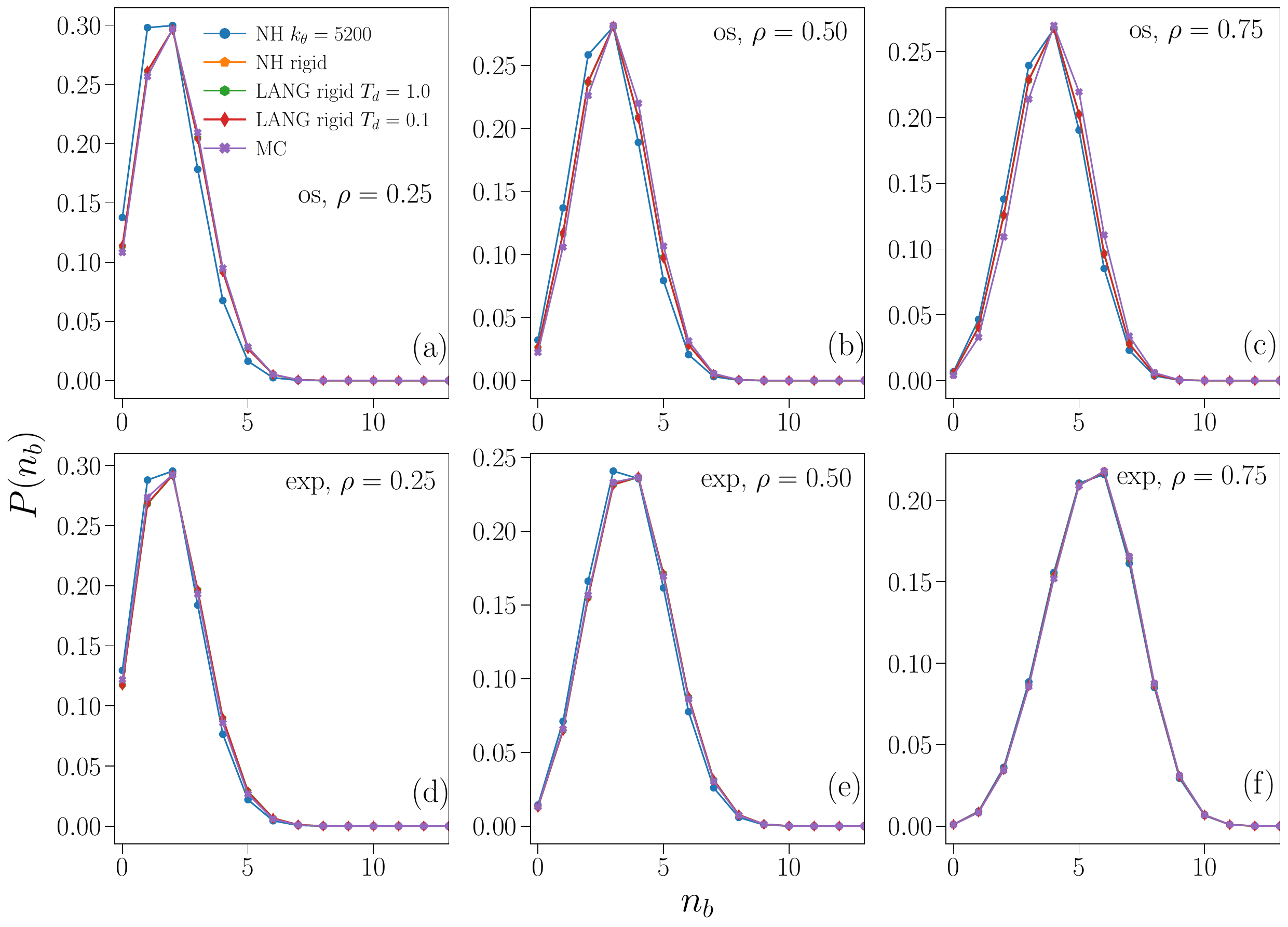}
\caption{Probability of the number of energetic bonds formed by a particle. 
(a)-(c): os model. (d)-(f): exp model.
(a) and (d): $\rho=0.25$.
(b) and (e): $\rho=0.50$. 
(c) and (f): $\rho=0.75$. 
Different colors and symbols are specified in the legend of the top left panel.
}
\label{fig:bonds} 
\end{center}      
\end{figure*}

We start by looking at the neighborhood of each particle, that we characterize via the number of pair configurations for which the potential energy is negative. We name such configurations ``energetic bonds''.
In Fig.~\ref{fig:bonds} we report the probability  of observing a certain number of energetic bonds per particle, obtained at different values of $\rho$, for both models, different simulations methods and different thermostats. We observe that the probability values obtained using different methods are compatible, within each model. The average number of energetic bonds consistently grows upon increasing $\rho$, as expected in a denser fluid. Further, the exp and os models display comparable distributions at $\rho=0.25, 0.5$; however, at $\rho=0.75$ the distribution for the exp model shows an overall shift to higher number of bonds, compared to the os case. Counter-intuitively, this is not matched by a more negative average potential energy per particle. Both effects are caused again by the longer interaction range of the exp model, as more same-charge contributions should be included for each particle. So, at the same time, the IPC fluid in the exp model is more bonded but on average each particle has a higher energy with respect to its os counterpart.   

\begin{figure*}[t]
\begin{center}
\includegraphics[width=\textwidth]{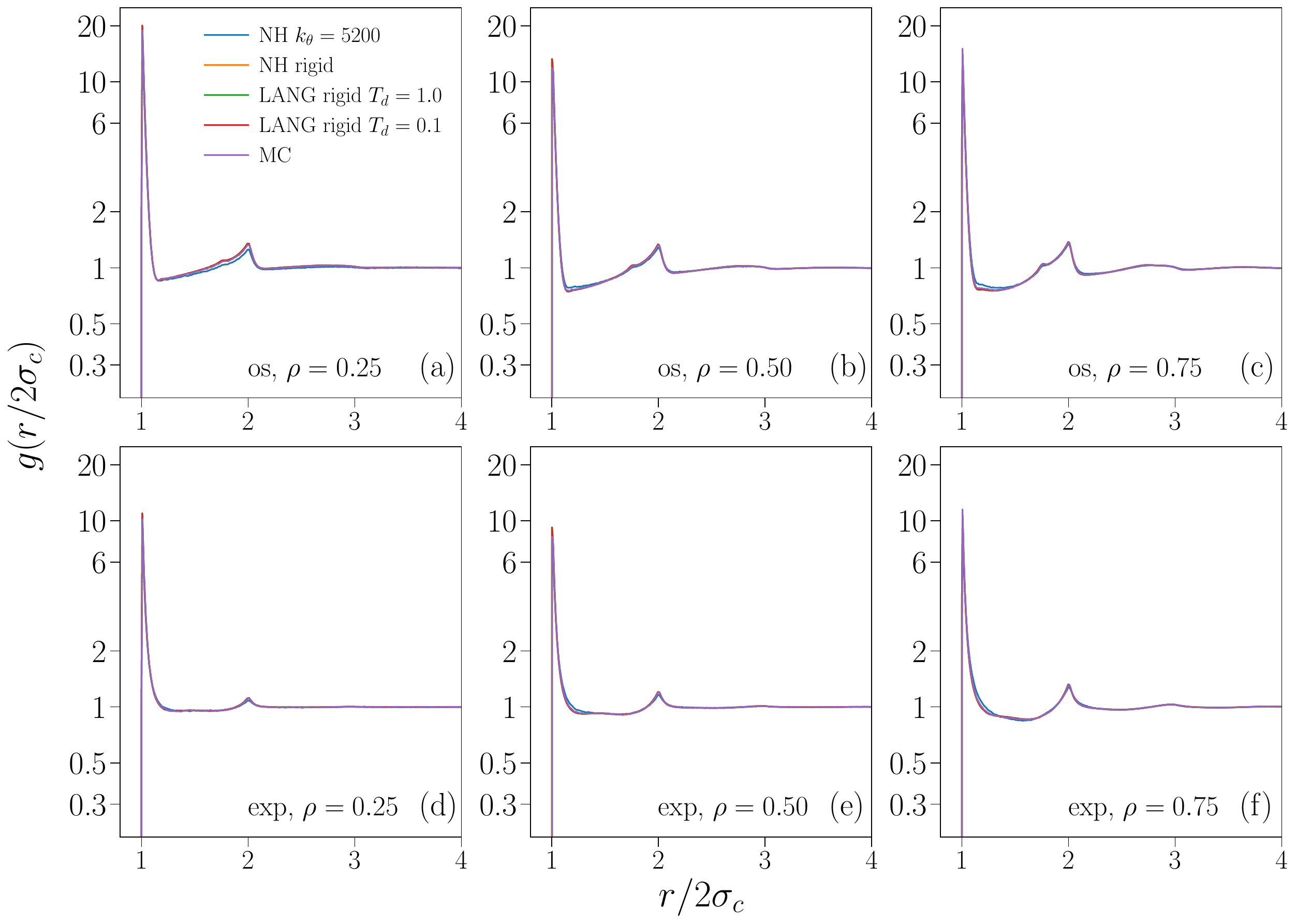}
\caption{Radial distribution function. 
(a)-(c): os model. (d)-(f): exp model.
(a) and (d): $\rho=0.25$.
(b) and (e): $\rho=0.50$. 
(c) and (f): $\rho=0.75$. 
 Different colors and symbols are specified in the legend of the top left panel.
}
\label{fig:rdf} 
\end{center}      
\end{figure*}

Finally, in Fig.~\ref{fig:rdf} we report the radial distribution functions $g(r)$ at different values of $\rho$, for both models, different simulations methods and different thermostats. Again, within each model, differences that arise from using different methods or thermostats are effectively negligible. On the other hand, it is interesting to notice that the $g(r)$ has slightly different signatures in the two models: in particular, the os model shows a more pronounced peak at $r=2\sigma=4\sigma_c$ signaling, overall, a more structured fluid. In general, at the temperature $T$ considered here, we observe a fluid state at all densities.

\subsection{Computational efficiency}

We report, in Table~\ref{tab:times}, the number of kilo-steps (ksteps i.e. $10^3$ steps) per second, averaged over time and over 8 parallel runs, with the corresponding standard deviation. Notice that, in the case of VMMC simulations, one MC step corresponds to $N$ trial moves. All the simulations have been performed on the same CPU (Intel Skylake Platinum 8174) on a single core. 

\begin{table*}[h]
\resizebox{\textwidth}{!}{
  \centering
  \begin{tabular}{c|cc|cc|cc}
    & \multicolumn{6}{c}{ksteps per second $(s^{-1})$} \\
    \hline
    & \multicolumn{2}{c|}{$\rho=0.25$} & \multicolumn{2}{c|}{$\rho=0.5$} & \multicolumn{2}{c}{$\rho=0.75$} \\
    \hline
        & os & exp & os & exp & os & exp \\
    \hline
MC & 0.23 & 0.18 & 0.15 & 0.11 & 0.09 & 0.07 \\
NH, RG & 2.68 $\pm$ 0.21 & 1.20 $\pm$ 0.04 & 2.09 $\pm$ 0.09 & 0.70 $\pm$ 0.01 & 1.69 $\pm$ 0.02 & 0.50 $\pm$ 0.02 \\
LG, RG & 2.62 $\pm$ 0.20 & 1.19 $\pm$ 0.05 & 2.06 $\pm$ 0.08 & 0.70 $\pm$ 0.01 & 1.68 $\pm$ 0.02 & 0.49 $\pm$ 0.02 \\
LG, RG & 2.62 $\pm$ 0.20 & 1.16 $\pm$ 0.03 & 2.06 $\pm$ 0.09 & 0.70 $\pm$ 0.01 & 1.68 $\pm$ 0.02 & 0.49 $\pm$ 0.03 \\
NH, $k=5.2 \cdot 10^3$ & 5.67 $\pm$ 0.60 & 1.62 $\pm$ 0.04 & 4.08 $\pm$ 0.25 & 0.89 $\pm$ 0.03 & 3.07 $\pm$ 0.05 & 0.61 $\pm$ 0.03 \\
\end{tabular}
}
  \caption{Average computational performance, measured in kilo-steps (ksteps) per second, of numerical simulations employing rigid methods for different models (os and exp) and different values of the density $\rho$.}
  \label{tab:times}
\end{table*}

We first compare, in Table~\ref{tab:times}, the ``rigid'' methods, i.e. the Monte Carlo and the constrained MD,  where the axiality and eccentricity of the IPCs are preserved by construction. We observe that the Monte Carlo code is one order of magnitude slower than the constrained MD: notice that both codes implement Verlet lists. Besides fine-scale optimisations, this performance is caused by two factors. The cluster nature of the algorithm requires, for every trial move, to build a cluster; this becomes expensive, especially at high density. Further, we perform $N$ cluster moves per step which, albeit limited to a maximum of $S^c$ recruited particles, are definitely more demanding than $N$ single particle moves or very few $\mathcal{O}(N)$ moves, as in more conventional cluster-based MC. However, we should also notice that cluster-based algorithms are often very efficient in producing decorrelated configurations; as mentioned, VMMC is well known for its ability to overcome kinetic barriers, especially at high density. It is also worth noting that the performance of the code further drops when comparing the two different models: the os model is systematically 30-40\% faster than the exp. The latter involves the evaluation of transcendental functions, that are computationally more expensive than the simple operations required by the former. However, this is clearly a second order effect, with respect to the overall computational complexity of the algorithm.\\
In contrast, when looking at the performances of the MD code, we notice that it is highly dependent on the chosen model, the os one being now significantly more efficient than the exp: indeed, simulations with the exp model take 2-3 times more time. This is entirely due, in the proposed LAMMPS implementation, to the longer range of the latter as, in both cases, we employ tabulated forces. Interestingly, the use of a different thermostat (NH or LG) mildly affects the results.\\When considering the chosen bead-spring implementation, performances increase considerably. Compared to the constrained-MD counterpart, the best improvements are still recorded for the os model, while the exp shows only a 30\% increase. 
Finally, as expected, the performance drops upon increasing the density.

\section{Conclusions}\label{sec:conclusions}

We have introduced a general model for simulating Inverse Patchy Colloids (IPCs) i.e. patchy particles featuring interaction that are inspired by heterogeneously charged systems. The model can indeed be used to describe specific physical systems and can be fitted to, e.g., a mean-field model~\cite{Bianchi2011} but it can be also used parametrically, as in~\cite{notarmuzishort,notarmuzilong} and in this paper.

In the model, an IPC is a collection of interacting sites with a specified geometrical arrangement: the interaction between the sites is characterized by a contact value and a geometrical weight that incorporates the dependence on the site-site distance. We showcase two IPC models: the overlap of spheres (os) and the exponential (exp). 
It is worth noting that the proposed framework is also able to describe conventional patchy particles. In fact, the characteristic energy values of the site-site interactions can be tuned to support repulsion as well as attraction between the different surface areas: when only attractive values are chosen, then the models represent conventional patchy colloids.

We showed that different simulation methods and different algorithms yield comparable results; the os model is, evidently, faster than the exp one and, thus, more suitable for studying generic properties, such as phase coexistence~\cite{notarmuzishort,notarmuzilong}. As mentioned, the bead-spring realization of the IPC has a lot of potential for further development, as it can accommodate for (and be fitted to) systems with moving patches~\cite{smallenburg2014erasing,smallenburg2015tuning,bianchi2015soft,rosales2020shape}. We proposed a set of parameters that, according to our tests, are suitable for efficient simulations of quasi-rigid IPC systems: however, we remark that other sets of parameters may be equally acceptable if, for example, a smaller value of $k_r$ is considered.

We remark that the model, being suitable for both MC and MD simulations, represents a versatile platform for simulations of colloids with heterogeneous directional interactions; its simple and relatively inexpensive nature allows, also by virtue of its implementation in LAMMPS, for simulations of large scale bulk systems. Moreover, the availability of the accompanying codes makes the model easily accessible for exploring a wide range of phenomena and facilitates straightforward extensions to systems with diverse charge surface patterns. 

Finally, we observe that the present investigations have been conducted in the fluid phase, where the different algorithms exhibit comparable results. However, it would be interesting to assess how these discrepancies evolve in more structured phases, such as crystalline or gel-like states, where the directional interactions play a more prominent role. These effects could offer deeper insights into the behavior of the different IPCs formulations in systems with higher order and might highlight the strengths and limitations of each algorithm in simulating such phases.

\section{Authors contributions}
All the authors were involved in the preparation of the manuscript.
All the authors have read and approved the final manuscript.

\section{Acknowledgments}
Financial support to carry this research was provided by the French Agency for Research (ANR) and by the Autrian Science Fund (FWF) under project numbers I-3577-N28 and Y-1163-N27. Computation time at the Vienna Scientific Cluster (VSC) is gratefully acknowledged.


\providecommand{\latin}[1]{#1}
\makeatletter
\providecommand{\doi}
  {\begingroup\let\do\@makeother\dospecials
  \catcode`\{=1 \catcode`\}=2 \doi@aux}
\providecommand{\doi@aux}[1]{\endgroup\texttt{#1}}
\makeatother
\providecommand*\mcitethebibliography{\thebibliography}
\csname @ifundefined\endcsname{endmcitethebibliography}
  {\let\endmcitethebibliography\endthebibliography}{}

\clearpage

\onecolumngrid
\begin{center}
\textbf{\Large  \\ \vspace*{1.5mm} -- Supporting Information -- \\ Simulating inverse patchy colloid models} \\
\vspace*{5mm}
Daniele Notarmuzi, Silvano Ferrari, Emanuele Locatelli, Emanuela Bianchi
\vspace*{10mm}
\end{center}





\section{Analytical details of the os model}

In the os model, the weight factors are defined as the normalized volumes of overlap between all the pairs of interaction spheres contributing to the specific $\alpha\beta$ interaction in the given AB configuration, they are thus expressed as the ratio between the $total$ overlap volume of the $\alpha$ and $\beta$ interaction spheres and a reference volume:
\begin{equation}
w_{\alpha\beta}^{\rm AB}=\frac{1}{V^{\rm Ref}}\sum_{r_{\alpha\beta}|{\rm AB}}V(r_{\alpha\beta})
\end{equation}
where the reference volume is the volume of the colloid 
\begin{equation}
V^{\rm Ref}= \frac{4}{3}\pi\sigma_c^3
\end{equation}
while the overlap volume between one $\alpha\beta$ pair is
\begin{equation}\label{eq:Overlap}
V(r_{\alpha\beta})=
\left\{
\begin{array}{rl}
 \frac{4}{3}\pi[{\rm min}(\sigma_{\alpha},\sigma_{\beta})]^3  &{\hspace{1em}\rm if\hspace{1em}} r_{\alpha\beta} \leq |\sigma_{\alpha}-\sigma_{\beta}|  \\
 v(r_{\alpha\beta}) &{\hspace{1em}\rm if\hspace{1em}} |\sigma_{\alpha}-\sigma_{\beta}| < r_{\alpha\beta} \leq \sigma_{\alpha}+\sigma_{\beta} \\
 0                       &{\hspace{1em}\rm if\hspace{1em}} r_{\alpha\beta} \geq \sigma_{\alpha}+\sigma_{\beta}
\end{array}
\right. 
\end{equation}
where $v(r_{\alpha\beta})$ is a simple algebraic expression
\begin{eqnarray}\label{eq:overlap}
v(r_{\alpha\beta})=&\frac{\pi}{3}\left[\left(2\sigma_{\alpha}+\frac{\sigma_{\alpha}^2-\sigma_{\beta}^2+r_{\alpha\beta}^2}{2r_{\alpha\beta}})(\sigma_{\alpha}-\frac{\sigma_{\alpha}^2-\sigma_{\beta}^2+r_{\alpha\beta}^2}{2r_{\alpha\beta}}\right)^2\right] \nonumber\\
+&\frac{\pi}{3}\left[\left(2\sigma_{\beta}-\frac{\sigma_{\alpha}^2-\sigma_{\beta}^2-r_{\alpha\beta}^2}{2r_{\alpha\beta}})(\sigma_{\beta}+\frac{\sigma_{\alpha}^2-\sigma_{\beta}^2-r_{\alpha\beta}^2}{2r_{\alpha\beta}}\right)^2\right].
\end{eqnarray}
%
%
%
 The assignment of the  $\epsilon_{\alpha\beta}$ is done by selecting reference pair configurations AB where the $\alpha\beta$ interaction type is the most relevant. Given two ts-IPCs, the configurations EE, EP and PP isolate the center-center, center-off-center and off-center-off-center interactions, respectively, if the following geometric conditions are satisfied: 
\begin{itemize}
\item in the EE configuration there must be no interaction (i) between the off-center sites, i.e., $\sigma_p \leq \sigma_c$, and (ii) between the center and the off-center site, i.e., $(\sigma_p + \sigma_c + \delta_c/2)^2 \leq a^2 + 4\sigma_c^2$
\item in the EP configuration there must be no interaction between the off-center sites, i.e., $4\sigma_p^2 \leq a^2 + (2\sigma_c - a)^2$
\end{itemize}
If these conditions are satisfied, the system of equations~(5) simplifies to 
\begin{equation}
  \begin{split}
    u^{\rm EE}    &= \epsilon_{\rm c,c} w^{\rm EE}_{\rm c,c} \\
    u^{\rm EP}    &= \epsilon_{\rm c,c} w^{\rm EP}_{\rm c,c} + \epsilon_{\rm c,oc} ws^{\rm EP}_{\rm c,oc} \\
    u^{\rm PP}    &= \epsilon_{\rm c,c} w^{\rm PP}_{\rm c,c} + \epsilon_{\rm c,oc} w^{\rm EP}_{\rm c,oc} + \epsilon_{\rm oc,oc} w^{\rm PP}_{\rm oc,oc} \\ 
   \end{split}.
\end{equation}

\section{Single particle properties}

In this section we report additional information on the single particle properties, i.e., distributions of the eccentricity (the distances between the central and the off-center charges) and of the axial angle (the angle between the thee charges in each IPC) for all systems studied  with the Nosé–Hoover thermostat. We consider all the simulations cases (i), (ii) and (iii) as in the main text: (i) fixing $k_r = k_{\theta}=k$, $10^3 \leq k \leq 10^4$ 
and $\Delta t=10^{-3}$, (ii) the same values of $k$ and variable $\Delta t$, specifically,
$\Delta t=10^{-3}$ for $k=10^3$,
$\Delta t=5 \cdot 10^{-4}$ for $2\cdot 10^3 \leq k
\leq 8 \cdot 10^3$ and $\Delta t=10^{-4}$ for $k=10^4$ and (iii) fixed  $\Delta t=10^{-3}$, $k_{r}=10^4$ and variable $100 \leq k_{\theta} \leq 10^4$.

\begin{figure}[h]
\begin{center}
\includegraphics[width=\textwidth]{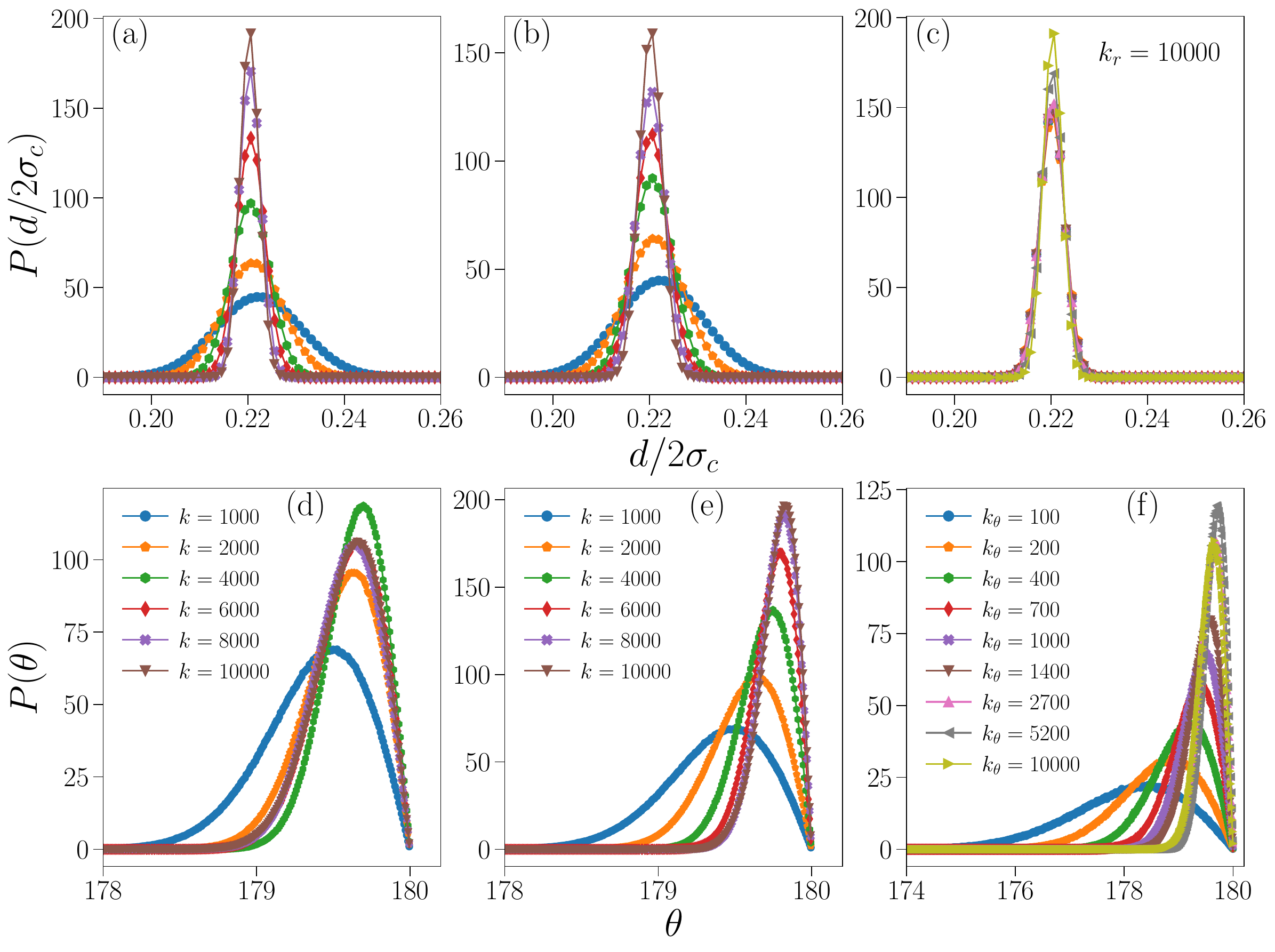}
\caption{Single particle properties for the os model at $\rho=0.25$ with the Nosé–Hoover thermostat. 
Top: eccentricity distribution.
Bottom: axial angle distribution.
(a),(d) Systems with $k_{\theta}=k_{r} \equiv k$ and $\Delta t=10^{-3}$.
(b),(e) Systems with $k_{\theta}=k_{r} \equiv k$ and variable $\Delta t$ (see Sec.~5 of the main text) 
(c),(f) Systems with $k_{r}=10^4$, $k_{\theta}$ as specified in the legend of panel (f) and $\Delta t=10^{-3}$.
}
\label{fig:SPos25} 
\end{center}      
\end{figure}

\begin{figure}[h]
\begin{center}
\includegraphics[width=\textwidth]{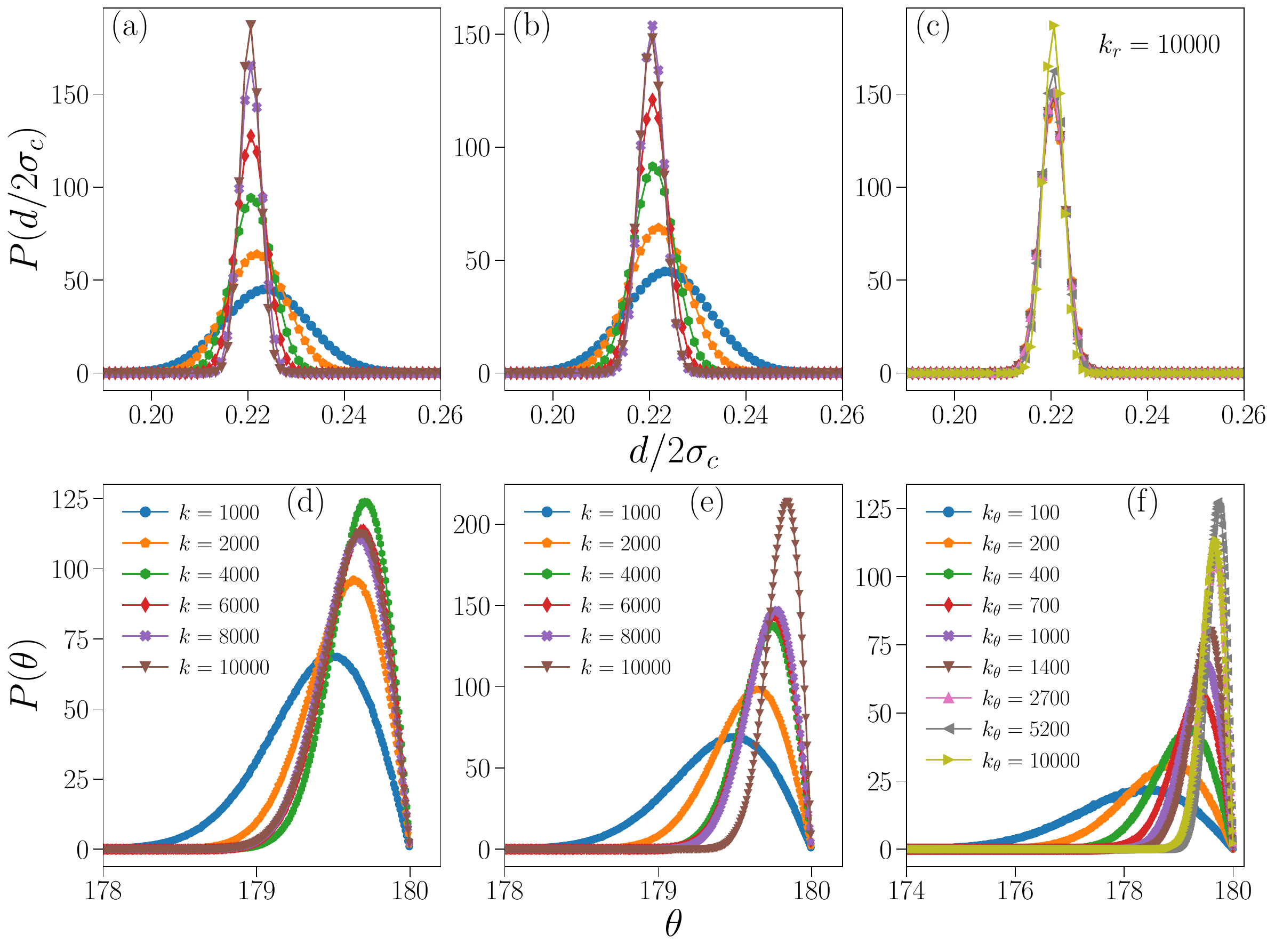}
\caption{Single particle properties for the os model at $\rho=0.75$ with the Nosé–Hoover thermostat. 
Top: eccentricity distribution.
Bottom: axial angle distribution.
(a),(d) Systems with $k_{\theta}=k_{r} \equiv k$ and $\Delta t=10^{-3}$.
(b),(e) Systems with $k_{\theta}=k_{r} \equiv k$ and variable $\Delta t$ (see Sec.~5 of the main text) 
(c),(f) Systems with $k_{r}=10^4$, $k_{\theta}$ as specified in the legend of panel (f) and $\Delta t=10^{-3}$.
}
\label{fig:SPos75} 
\end{center}      
\end{figure}

\begin{figure}[h]
\begin{center}
\includegraphics[width=\textwidth]{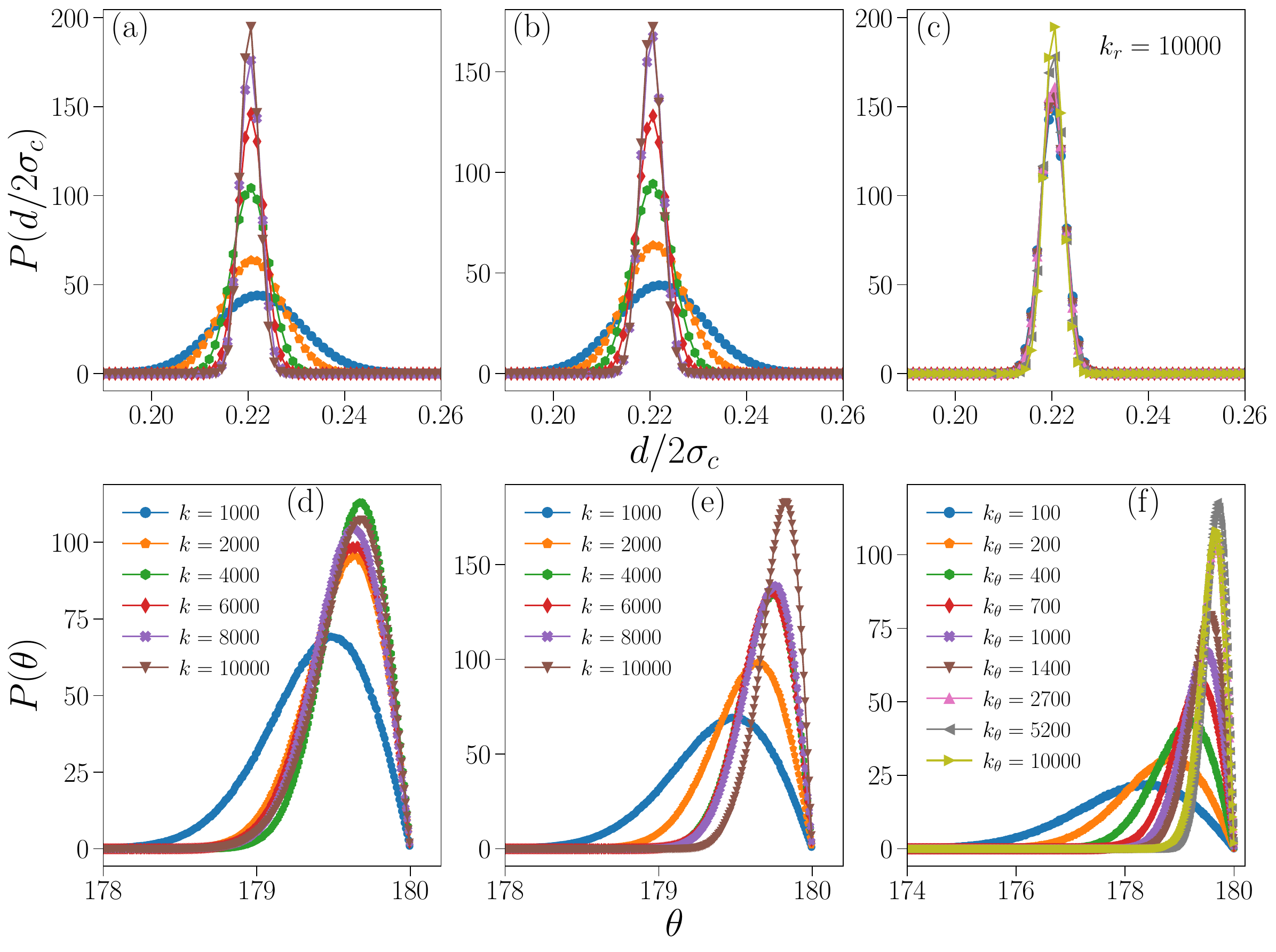}
\caption{Single particle properties for the exp model at $\rho=0.25$ with the Nosé–Hoover thermostat. 
Top: eccentricity distribution.
Bottom: axial angle distribution.
(a),(d) Systems with $k_{\theta}=k_{r} \equiv k$ and $\Delta t=10^{-3}$.
(b),(e) Systems with $k_{\theta}=k_{r} \equiv k$ and variable $\Delta t$ (see Sec.~5 of the main text) 
(c),(f) Systems with $k_{r}=10^4$, $k_{\theta}$ as specified in the legend of panel (f) and $\Delta t=10^{-3}$.
}
\label{fig:SPexp25} 
\end{center}      
\end{figure}

\begin{figure}[h]
\begin{center}
\includegraphics[width=\textwidth]{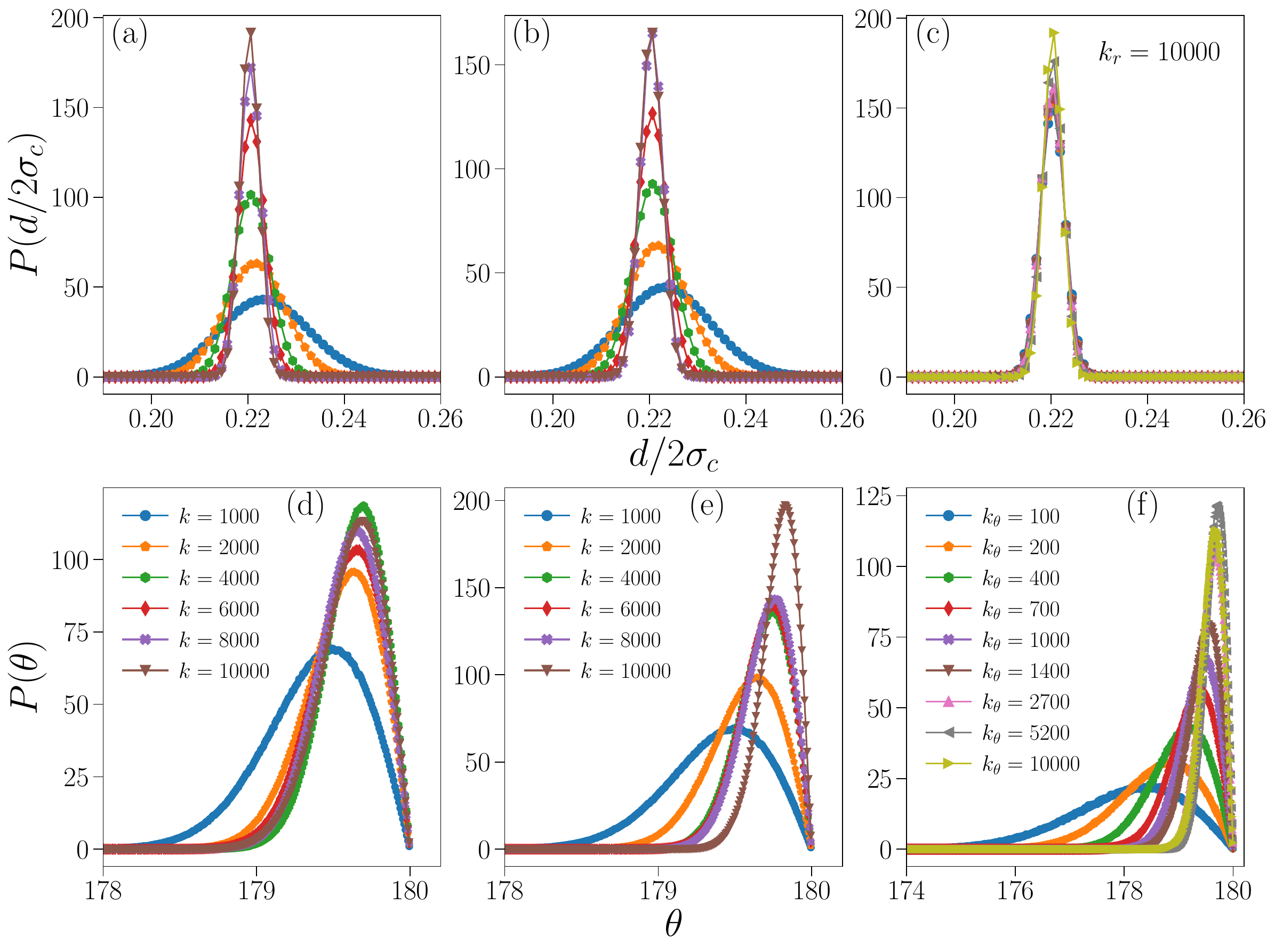}
\caption{Single particle properties for the exp model at $\rho=0.50$ with the Nosé–Hoover thermostat. 
Top: eccentricity distribution.
Bottom: axial angle distribution.
(a),(d) Systems with $k_{\theta}=k_{r} \equiv k$ and $\Delta t=10^{-3}$.
(b),(e) Systems with $k_{\theta}=k_{r} \equiv k$ and variable $\Delta t$ (see Sec.~5 of the main text) 
(c),(f) Systems with $k_{r}=10^4$, $k_{\theta}$ as specified in the legend of panel (f) and $\Delta t=10^{-3}$.
}
\label{fig:SPexp5} 
\end{center}      
\end{figure}

We show results for the os model at $\rho=0.25$ in Fig.~\ref{fig:SPos25} and at $\rho=0.75$ in Fig.~\ref{fig:SPos75} as well as 
results for the exp model at $\rho=0.25$ in Fig.~\ref{fig:SPexp25}, $\rho=0.50$ in Fig.~\ref{fig:SPexp5} and $\rho=0.75$ in Fig.~\ref{fig:SPos75}. Note that results for the os model at $\rho=0.50$ are shown in Fig.~3 of the main paper. 
The figures show that the conclusions drawn in the main paper regarding the interplay between $k_{r}$, $k_{\theta}$ and
$\Delta t$ hold at any density and for both the os and the exp model.

\begin{figure}[h]
\begin{center}
\includegraphics[width=\textwidth]{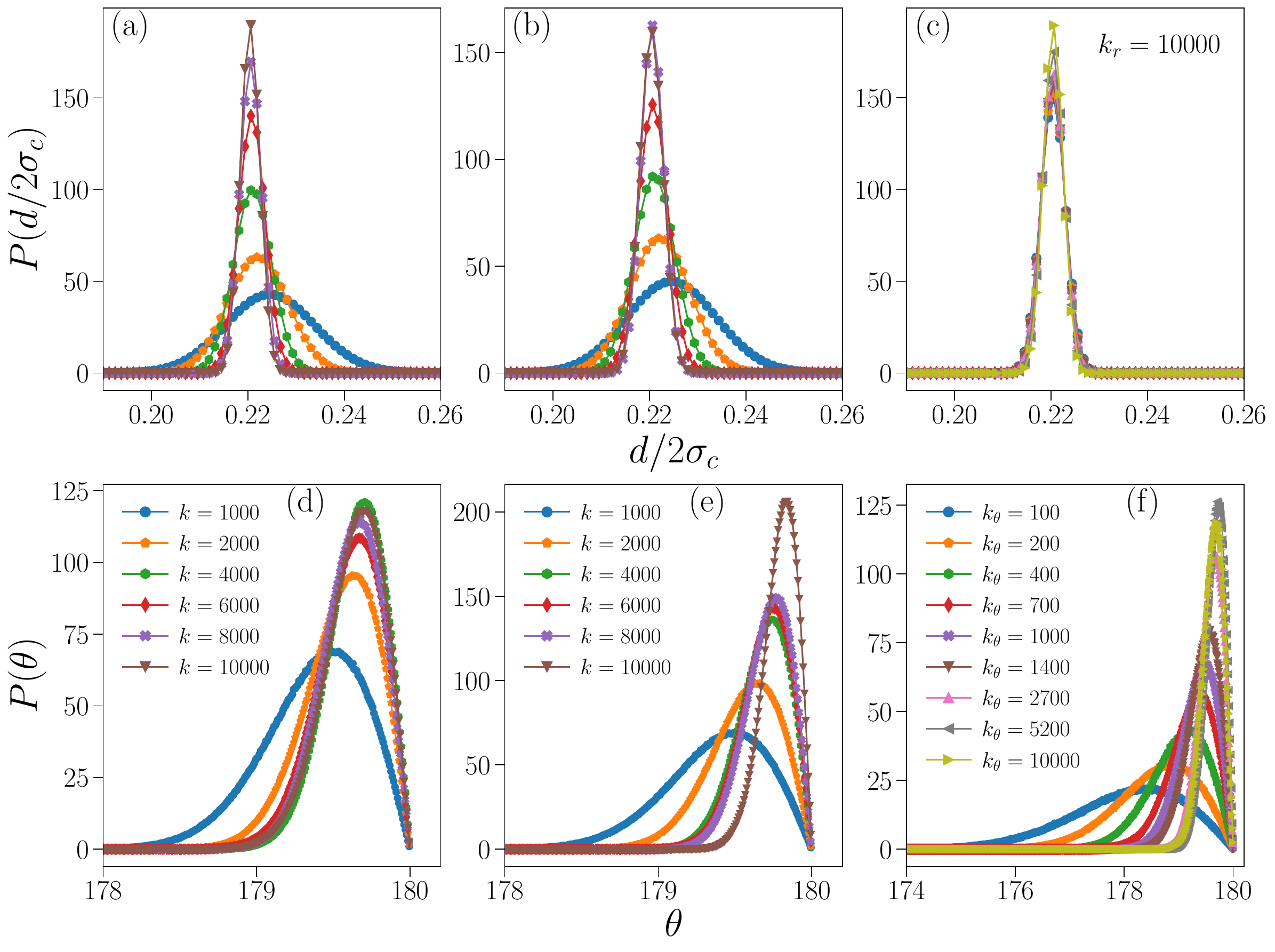}
\caption{Single particle properties for the exp model at $\rho=0.75$ with the Nosé–Hoover thermostat. 
Top: eccentricity distribution.
Bottom: axial angle distribution.
(a),(d) Systems with $k_{\theta}=k_{r} \equiv k$ and $\Delta t=10^{-3}$.
(b),(e) Systems with $k_{\theta}=k_{r} \equiv k$ and variable $\Delta t$ (see Sec.~5 of the main text) 
(c),(f) Systems with $k_{r}=10^4$, $k_{\theta}$ as specified in the legend of panel (f) and $\Delta t=10^{-3}$.
}
\label{fig:SPexp75} 
\end{center}      
\end{figure}

\section{Fluid structure properties}

In this section we provide additional information on the structure properties of the fluid. In particular, we look at the probability of having a certain number of ``energetic bonds'' in the neighborhood of a particle and at the pair distribution function. We look here more in detail at the cases (i), (ii) and (iii), detailed above, for the os and exp model.

\begin{figure}[h]
\begin{center}
\includegraphics[width=\textwidth]{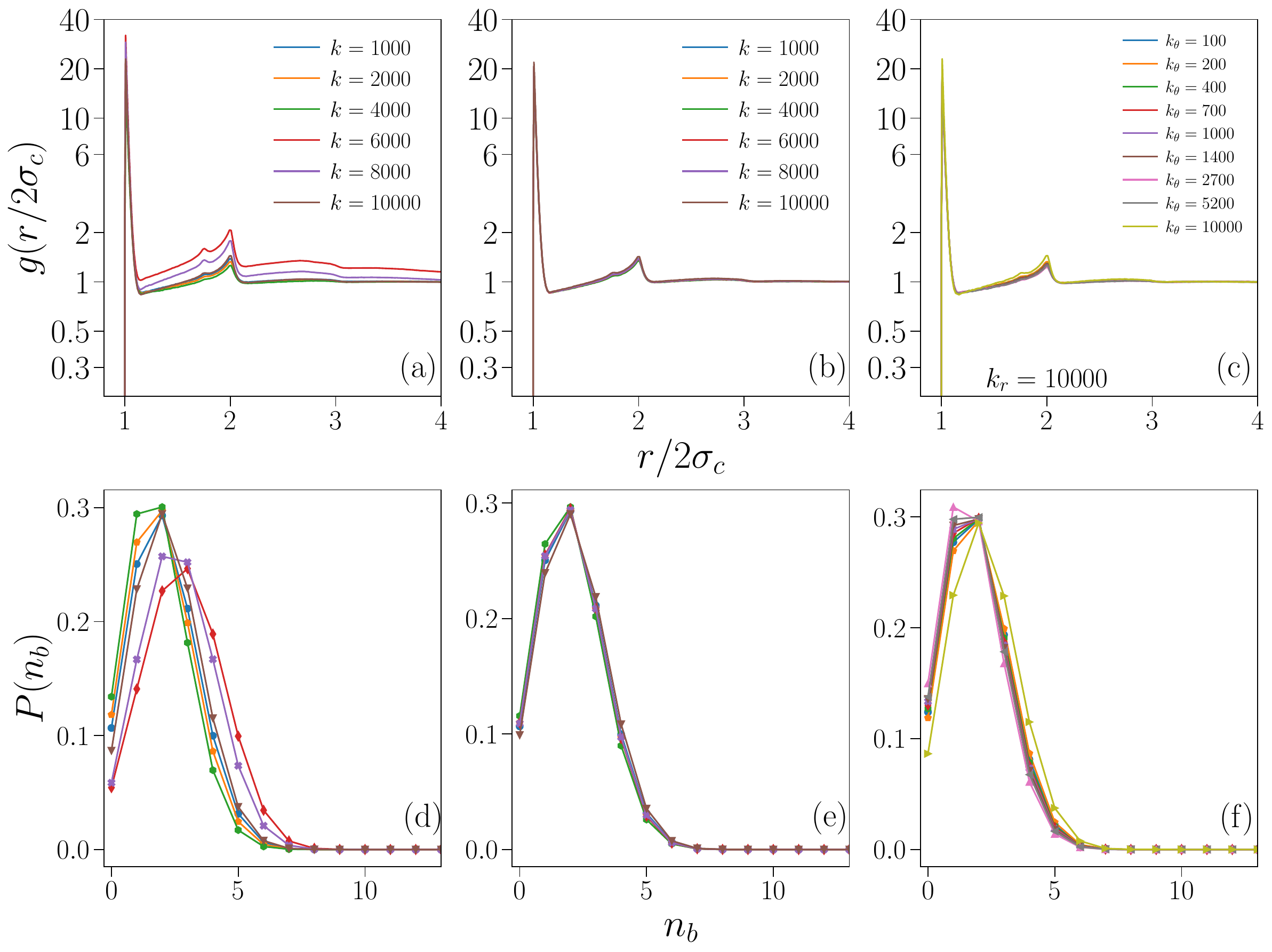}
\caption{Pair properties for the os model at $\rho=0.25$ with the Nosé–Hoover thermostat. 
Top: radial distribution function.
Bottom: probability of the number of energetic bonds.
(a),(d) Systems with $k_{\theta}=k_{r} \equiv k$ and $\Delta t=10^{-3}$.
(b),(e) Systems with $k_{\theta}=k_{r} \equiv k$ and variable $\Delta t$ (see Sec.~5 of the main text) 
(c),(f) Systems with $k_{r}=10^4$, $k_{\theta}$ as specified in the legend of panel (f) and $\Delta t=10^{-3}$.
}
\label{fig:Pos25} 
\end{center}      
\end{figure}

\begin{figure}[h]
\begin{center}
\includegraphics[width=\textwidth]{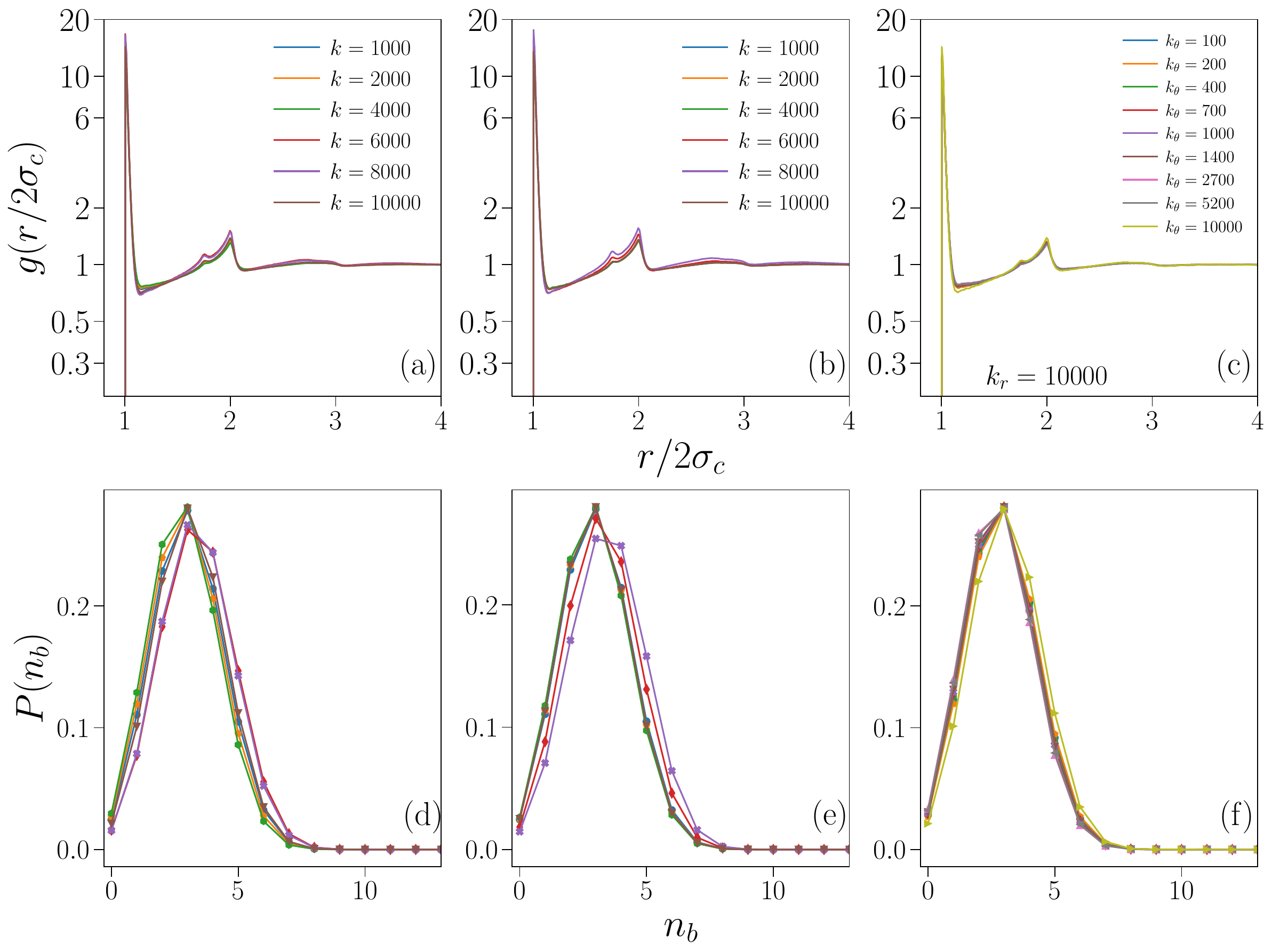}
\caption{Pair properties for the os model at $\rho=0.50$ with the Nosé–Hoover thermostat. 
Top: radial distribution function.
Bottom: probability of the number of energetic bonds.
(a),(d) Systems with $k_{\theta}=k_{r} \equiv k$ and $\Delta t=10^{-3}$.
(b),(e) Systems with $k_{\theta}=k_{r} \equiv k$ and variable $\Delta t$ (see Sec.~5 of the main text) 
(c),(f) Systems with $k_{r}=10^4$, $k_{\theta}$ as specified in the legend of panel (f) and $\Delta t=10^{-3}$.
}
\label{fig:Pos5} 
\end{center}      
\end{figure}

\begin{figure}[h]
\begin{center}
\includegraphics[width=\textwidth]{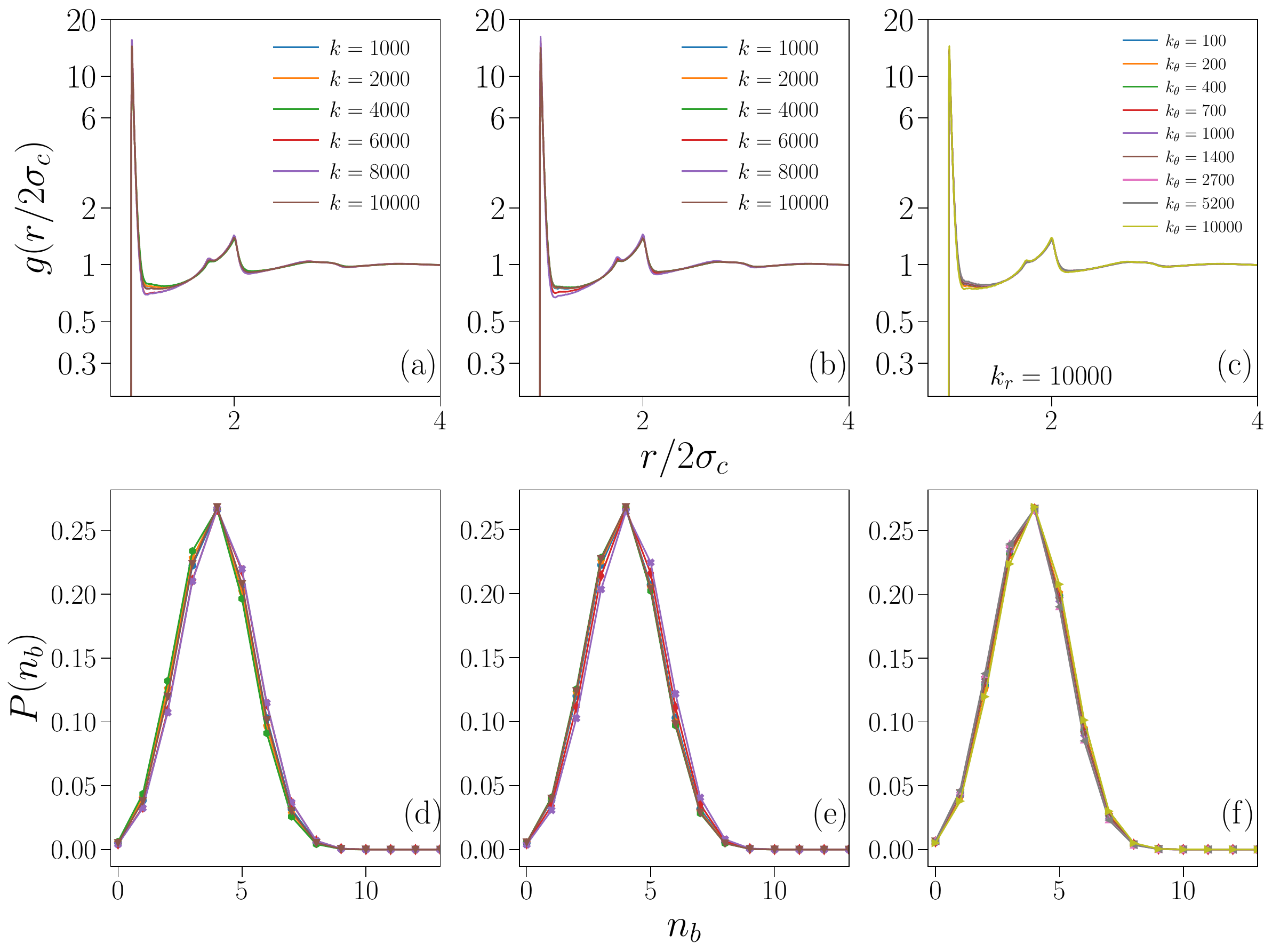}
\caption{Pair properties for the os model at $\rho=0.75$ with the Nosé–Hoover thermostat. 
Top: radial distribution function.
Bottom: probability of the number of energetic bonds.
(a),(d) Systems with $k_{\theta}=k_{r} \equiv k$ and $\Delta t=10^{-3}$.
(b),(e) Systems with $k_{\theta}=k_{r} \equiv k$ and variable $\Delta t$ (see Sec.~5 of the main text) 
(c),(f) Systems with $k_{r}=10^4$, $k_{\theta}$ as specified in the legend of panel (f) and $\Delta t=10^{-3}$.
}
\label{fig:Pos75} 
\end{center}      
\end{figure}

\begin{figure}[h]
\begin{center}
\includegraphics[width=\textwidth]{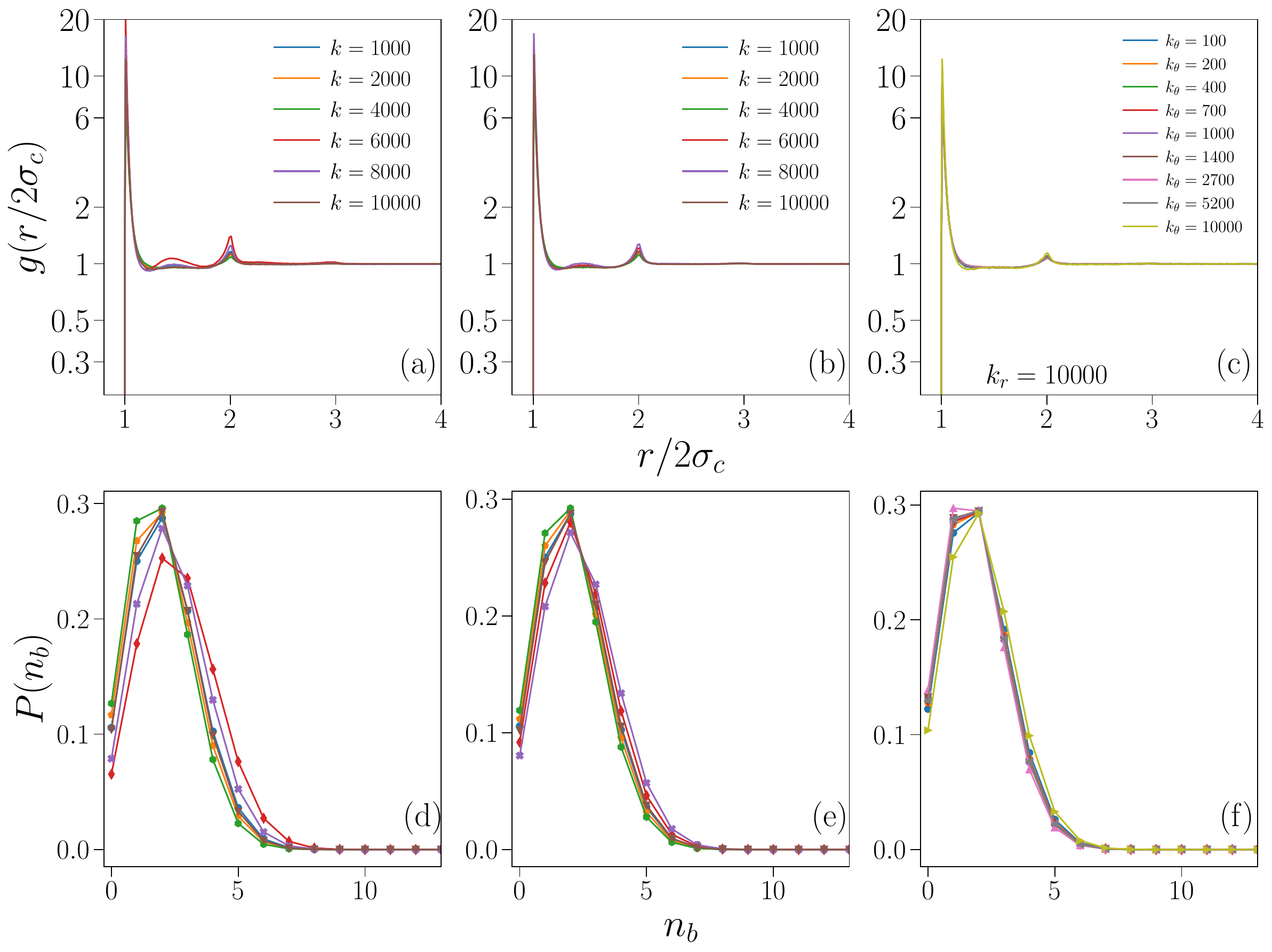}
\caption{Pair properties for the exp model at $\rho=0.25$ with the Nosé–Hoover thermostat. 
Top: radial distribution function.
Bottom: probability of the number of energetic bonds.
(a),(d) Systems with $k_{\theta}=k_{r} \equiv k$ and $\Delta t=10^{-3}$.
(b),(e) Systems with $k_{\theta}=k_{r} \equiv k$ and variable $\Delta t$ (see Sec.~5 of the main text) 
(c),(f) Systems with $k_{r}=10^4$, $k_{\theta}$ as specified in the legend of panel (f) and $\Delta t=10^{-3}$.
}
\label{fig:Pexp25} 
\end{center}      
\end{figure}

\begin{figure}[h]
\begin{center}
\includegraphics[width=\textwidth]{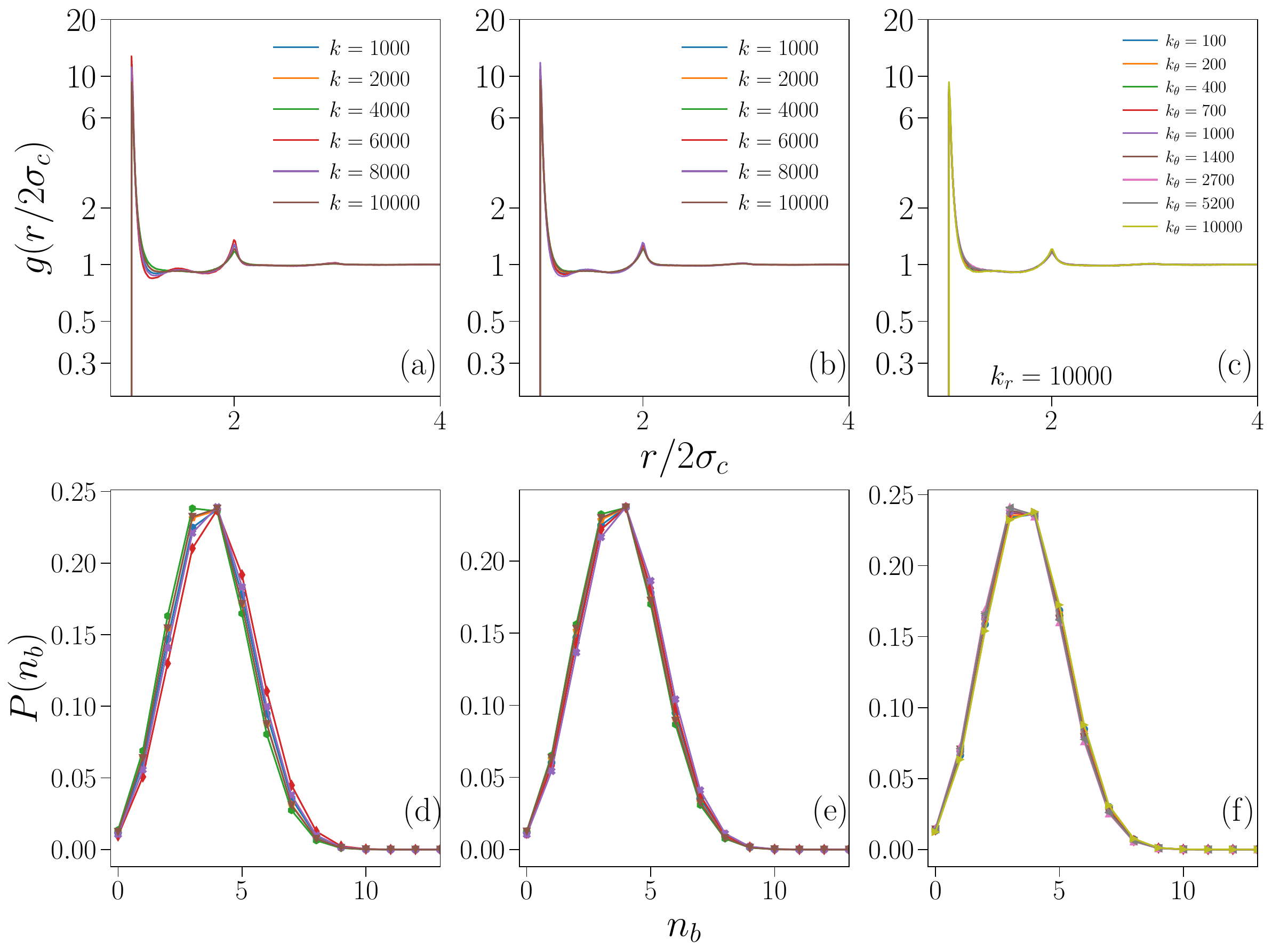}
\caption{Pair properties for the exp model at $\rho=0.50$ with the Nosé–Hoover thermostat. 
Top: radial distribution function.
Bottom: probability of the number of energetic bonds.
(a),(d) Systems with $k_{\theta}=k_{r} \equiv k$ and $\Delta t=10^{-3}$.
(b),(e) Systems with $k_{\theta}=k_{r} \equiv k$ and variable $\Delta t$ (see Sec.~5 of the main text) 
(c),(f) Systems with $k_{r}=10^4$, $k_{\theta}$ as specified in the legend of panel (f) and $\Delta t=10^{-3}$.
}
\label{fig:Pexp5} 
\end{center}      
\end{figure}

Specifically, we show results for the os model at $\rho=0.25$ in Fig.~\ref{fig:Pos25}, $\rho=0.5$ in Fig.~\ref{fig:Pos5}, $\rho=0.75$ in Fig.~\ref{fig:SPos75} as well as 
results for the exp model at $\rho=0.25$ in Fig.~\ref{fig:Pexp25}, $\rho=0.50$ in Fig.~\ref{fig:Pexp5} and $\rho=0.75$ in Fig.~\ref{fig:Pos75}.

\begin{figure}[h]
\begin{center}
\includegraphics[width=\textwidth]{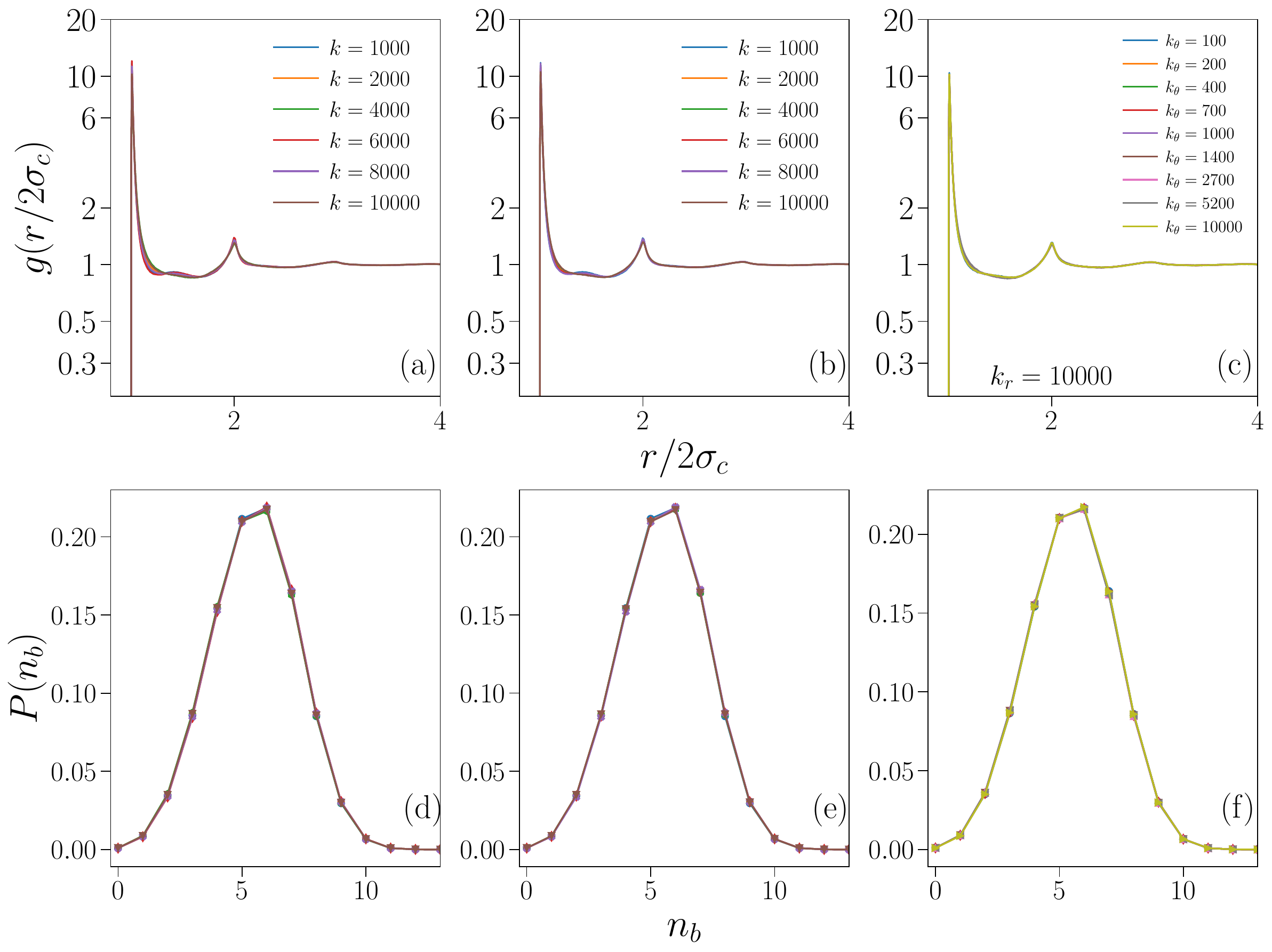}
\caption{Pair properties for the exp model at $\rho=0.75$ with the Nosé–Hoover thermostat. 
Top: radial distribution function.
Bottom: probability of the number of energetic bonds.
(a),(d) Systems with $k_{\theta}=k_{r} \equiv k$ and $\Delta t=10^{-3}$.
(b),(e) Systems with $k_{\theta}=k_{r} \equiv k$ and variable $\Delta t$ (see Sec.~5 of the main text) 
(c),(f) Systems with $k_{r}=10^4$, $k_{\theta}$ as specified in the legend of panel (f) and $\Delta t=10^{-3}$.
}
\label{fig:Pexp75} 
\end{center}      
\end{figure}

The results reported show that, indeed, the systems that we discard on the basis of the single particle properties do show differences in both the energetic bonds and radial distribution functions. As mentioned, these differences are mitigated by decreasing $\Delta t$ at the expense of the computational efficiency. Decoupling $k_{\theta}$ from $k_r$ solves both issues. Similar considerations hold for both os and exp models. Notice that for the exp model at $\rho=0.75$ (Fig.~\ref{fig:Pexp75}) all the different choices of parameters lead to very comparable results. This indeed highlights two trends that are visible throughout the data: the exp model is less sensible to the choice of parameters than the os model and, further, at high density the model and the implementation details are also less important.

\section{Thermodynamics}

In this section we provide additional information on the thermodynamics. In particular, we look at the average kinetic temperature and at the pair energy per particle. Again, we look here more in detail at the cases (i), (ii) and (iii), detailed above, for the os and exp model. We also include comparison with the rigid body (constrained MD) simulations.

The thermodynamic quantities, at every value of $\rho$, show that selecting the wrong parameters in the bead-spring model leads to considerable discrepancies in the measured pair energy. However, as shown in the main text, matching the pair energy is not the only measure for the choice of the parameters: single-particle statistics, as well as computational efficiency, should also be accounted. 

\begin{table*}[h]
\resizebox{\textwidth}{!}{
  \centering
  \begin{tabular}{c|cc|cc}
    & \multicolumn{2}{c|}{$T$} & \multicolumn{2}{c}{$U$} \\
    \hline
                                & os & exp & os & exp \\
    \hline
NH, RG, $T_d=0.10$ & 0.1500 $\pm$ 0.0029 & 0.1500 $\pm$ 0.0030 & -0.6346 $\pm$ 0.0188 & -0.3118 $\pm$ 0.0121 \\
LG, RG, $T_d=1.00$ & 0.1501 $\pm$ 0.0030 & 0.1501 $\pm$ 0.0030 & -0.6339 $\pm$ 0.0179 & -0.3120 $\pm$ 0.0119 \\
LG, RG, $T_d=0.10$ & 0.1503 $\pm$ 0.0037 & 0.1502 $\pm$ 0.0035 & -0.6315 $\pm$ 0.0242 & -0.3114 $\pm$ 0.0135 \\
\hline
    \multicolumn{5}{c}{(i)} \\
\hline
NH, $k=1 \cdot 10^3$, $T_d=0.10$ & 0.1500 $\pm$ 0.0022 & 0.1500 $\pm$ 0.0022 & -0.6861 $\pm$ 0.0197 & -0.4063 $\pm$ 0.0149 \\
NH, $k=2 \cdot 10^3$, $T_d=0.10$ & 0.1500 $\pm$ 0.0022 & 0.1500 $\pm$ 0.0022 & -0.6279 $\pm$ 0.0184 & -0.3325 $\pm$ 0.0126 \\
NH, $k=4 \cdot 10^3$, $T_d=0.10$ & 0.1500 $\pm$ 0.0022 & 0.1500 $\pm$ 0.0023 & -0.5686 $\pm$ 0.0162 & -0.2837 $\pm$ 0.0114 \\
NH, $k=6 \cdot 10^3$, $T_d=0.10$ & 0.1500 $\pm$ 0.0023 & 0.1501 $\pm$ 0.0024 & -0.9558 $\pm$ 0.0379 & -0.5643 $\pm$ 0.0228 \\
NH, $k=8 \cdot 10^3$, $T_d=0.10$ & 0.1501 $\pm$ 0.0024 & 0.1500 $\pm$ 0.0024 & -0.8885 $\pm$ 0.0263 & -0.4696 $\pm$ 0.0163 \\
NH, $k=10 \cdot 10^3$, $T_d=0.10$ & 0.1500 $\pm$ 0.0023 & 0.1500 $\pm$ 0.0024 & -0.7249 $\pm$ 0.0210 & -0.3581 $\pm$ 0.0197 \\
    \hline
    \multicolumn{5}{c}{(ii)} \\
    \hline
NH, $k=1 \cdot 10^3$, $T_d=0.10$ & 0.1500 $\pm$ 0.0022 & 0.1500 $\pm$ 0.0022 & -0.6861 $\pm$ 0.0197 & -0.4063 $\pm$ 0.0149 \\
NH, $k=2 \cdot 10^3$, $T_d=0.05$ & 0.1500 $\pm$ 0.0022 & 0.1500 $\pm$ 0.0022 & -0.6608 $\pm$ 0.0193 & -0.3525 $\pm$ 0.0132 \\
NH, $k=4 \cdot 10^3$, $T_d=0.05$ & 0.1500 $\pm$ 0.0022 & 0.1500 $\pm$ 0.0022 & -0.6326 $\pm$ 0.0186 & -0.3145 $\pm$ 0.0121 \\
NH, $k=6 \cdot 10^3$, $T_d=0.01$ & 0.1500 $\pm$ 0.0022 &  & -0.6384 $\pm$ 0.0456 &  \\
NH, $k=6 \cdot 10^3$, $T_d=0.05$ &  & 0.1500 $\pm$ 0.0023 & & -0.4222 $\pm$ 0.0167 \\
NH, $k=8 \cdot 10^3$, $T_d=0.01$ & 0.1500 $\pm$ 0.0022 &  & -0.6352 $\pm$ 0.0476 &  \\
NH, $k=8 \cdot 10^3$, $T_d=0.05$ &  & 0.1500 $\pm$ 0.0024 &  & -0.4786 $\pm$ 0.0215 \\
NH, $k=10 \cdot 10^3$, $T_d=0.01$ & 0.1500 $\pm$ 0.0023 & 0.1500 $\pm$ 0.0023 & -0.6768 $\pm$ 0.0465 & -0.3689 $\pm$ 0.0225 \\
\hline
    \multicolumn{5}{c}{(iii)} \\
    \hline
NH, $k_{\theta}= 10^2$, $T_d=0.10$ & 0.1500 $\pm$ 0.0022 & 0.1500 $\pm$ 0.0022 & -0.6007 $\pm$ 0.0169 & -0.2974 $\pm$ 0.0116 \\
NH, $k_{\theta}=2 \cdot 10^2$, $T_d=0.10$ & 0.1500 $\pm$ 0.0022 & 0.1500 $\pm$ 0.0022 & -0.6180 $\pm$ 0.0181 & -0.2839 $\pm$ 0.0132 \\
NH, $k_{\theta}=4 \cdot 10^2$, $T_d=0.10$ & 0.1500 $\pm$ 0.0022 & 0.1500 $\pm$ 0.0022 & -0.5901 $\pm$ 0.0165 & -0.2785 $\pm$ 0.0135 \\
NH, $k_{\theta}=7 \cdot 10^2$, $T_d=0.10$ & 0.1500 $\pm$ 0.0022 & 0.1500 $\pm$ 0.0022 & -0.5816 $\pm$ 0.0172 & -0.2773 $\pm$ 0.0127 \\
NH, $k_{\theta}= 10^3$, $T_d=0.10$ & 0.1500 $\pm$ 0.0022 & 0.1500 $\pm$ 0.0023 & -0.5737 $\pm$ 0.0163 & -0.2734 $\pm$ 0.0129 \\
NH, $k_{\theta}=1.4 \cdot 10^3$, $T_d=0.10$ & 0.1500 $\pm$ 0.0022 & 0.1500 $\pm$ 0.0022 & -0.5661 $\pm$ 0.0163 & -0.2690 $\pm$ 0.0126 \\
NH, $k_{\theta}=2.7 \cdot 10^3$, $T_d=0.10$ & 0.1500 $\pm$ 0.0022 & 0.1500 $\pm$ 0.0022 & -0.5285 $\pm$ 0.0161 & -0.2490 $\pm$ 0.0130 \\
NH, $k_{\theta}=5.2 \cdot 10^3$, $T_d=0.10$ & 0.1500 $\pm$ 0.0023 & 0.1500 $\pm$ 0.0023 & -0.5570 $\pm$ 0.0162 & -0.2737 $\pm$ 0.0153 \\
NH, $k_{\theta}= 10^4$, $T_d=0.10$ & 0.1501 $\pm$ 0.0024 & 0.1500 $\pm$ 0.0024 & -0.7237 $\pm$ 0.0209 & -0.3582 $\pm$ 0.0195 \\
\hline
LG, $k=10 \cdot 10^3$, $T_d = 1.00$ & 0.1509 $\pm$ 0.0023 & 0.1509 $\pm$ 0.0022 & -0.6401 $\pm$ 0.0179 & -0.3200 $\pm$ 0.0120 \\
LG, $k=10 \cdot 10^3$, $T_d = 0.10$ & 0.1701 $\pm$ 0.0032 & 0.1670 $\pm$ 0.0031 & -0.5337 $\pm$ 0.0190 & -0.2809 $\pm$ 0.0119 \\
\end{tabular}
}
  \caption{Average kinetic temperature and pair energy per particle in LAMMPS simulations at $\rho=0.25$. Groups (i), (ii) and (iii) as in main text. Datasets with varying $k_{\theta}$
  are generated using $k_{r}=10^4$, datasets with a unique value of $k$ are generated using $k_{\theta}=k_{r}$.}
  \label{tab:TU_0.25}
\end{table*}


\begin{table*}[h]
\resizebox{\textwidth}{!}{
  \centering
  \begin{tabular}{c|cc|cc}
    & \multicolumn{2}{c|}{$T$} & \multicolumn{2}{c}{$U$} \\
    \hline
             & os & exp & os & exp \\
    \hline
NH, RG, $T_d=0.10$ & 0.1500 $\pm$ 0.0030 & 0.1501 $\pm$ 0.0030 & -0.9370 $\pm$ 0.0156 & -0.5768 $\pm$ 0.0132 \\
LG, RG, $T_d=1.00$ & 0.1502 $\pm$ 0.0030 & 0.1501 $\pm$ 0.0031 & -0.9362 $\pm$ 0.0161 & -0.5765 $\pm$ 0.0132 \\
LG, RG, $T_d=0.10$ & 0.1505 $\pm$ 0.0039 & 0.1504 $\pm$ 0.0037 & -0.9332 $\pm$ 0.0234 & -0.5755 $\pm$ 0.0167 \\
\hline
    \multicolumn{5}{c}{(i)} \\
\hline
NH, $k=1 \cdot 10^3$, $T_d=0.10$ & 0.1500 $\pm$ 0.0022 & 0.1500 $\pm$ 0.0022 & -1.0030 $\pm$ 0.0166 & -0.7275 $\pm$ 0.0164 \\
NH, $k=2 \cdot 10^3$, $T_d=0.10$ & 0.1501 $\pm$ 0.0023 & 0.1500 $\pm$ 0.0022 & -0.9448 $\pm$ 0.0160 & -0.6190 $\pm$ 0.0143 \\
NH, $k=4 \cdot 10^3$, $T_d=0.10$ & 0.1500 $\pm$ 0.0022 & 0.1500 $\pm$ 0.0022 & -0.8960 $\pm$ 0.0151 & -0.5411 $\pm$ 0.0130 \\
NH, $k=6 \cdot 10^3$, $T_d=0.10$ & 0.1501 $\pm$ 0.0023 & 0.1501 $\pm$ 0.0023 & -1.1297 $\pm$ 0.0214 & -0.7732 $\pm$ 0.0171 \\
NH, $k=8 \cdot 10^3$, $T_d=0.10$ & 0.1501 $\pm$ 0.0023 & 0.1501 $\pm$ 0.0024 & -1.1232 $\pm$ 0.0180 & -0.6961 $\pm$ 0.0158 \\
NH, $k=10 \cdot 10^3$, $T_d=0.10$ & 0.1501 $\pm$ 0.0023 & 0.1501 $\pm$ 0.0023 & -1.0082 $\pm$ 0.0167 & -0.5945 $\pm$ 0.0190 \\
\hline
    \multicolumn{5}{c}{(ii)} \\
\hline
NH, $k=1 \cdot 10^3$, $T_d=0.10$ & 0.1500 $\pm$ 0.0022 & 0.1500 $\pm$ 0.0022 & -1.0030 $\pm$ 0.0166 & -0.7275 $\pm$ 0.0164 \\
NH, $k=2 \cdot 10^3$, $T_d=0.05$ & 0.1500 $\pm$ 0.0022 & 0.1500 $\pm$ 0.0022 & -0.9702 $\pm$ 0.0161 & -0.6420 $\pm$ 0.0144 \\
NH, $k=4 \cdot 10^3$, $T_d=0.05$ & 0.1500 $\pm$ 0.0022 & 0.1500 $\pm$ 0.0022 & -0.9418 $\pm$ 0.0156 & -0.5890 $\pm$ 0.0132 \\
NH, $k=6 \cdot 10^3$, $T_d=0.05$ & 0.1500 $\pm$ 0.0023 & 0.1500 $\pm$ 0.0023 & -1.0733 $\pm$ 0.0198 & -0.6754 $\pm$ 0.0147 \\
NH, $k=8 \cdot 10^3$, $T_d=0.05$ & 0.1500 $\pm$ 0.0023 & 0.1500 $\pm$ 0.0024 & -1.1662 $\pm$ 0.0234 & -0.7208 $\pm$ 0.0179 \\
NH, $k=10 \cdot 10^3$, $T_d=0.01$ & 0.1500 $\pm$ 0.0023 & 0.1500 $\pm$ 0.0023 & -0.9478 $\pm$ 0.0311 & -0.5959 $\pm$ 0.0178 \\
\hline
    \multicolumn{5}{c}{(iii)} \\
\hline
NH, $k_{\theta}=10^2$, $T_d=0.10$ & 0.1501 $\pm$ 0.0022 & 0.1500 $\pm$ 0.0022 & -0.9130 $\pm$ 0.0152 & -0.5613 $\pm$ 0.0129 \\
NH, $k_{\theta}=2 \cdot 10^2$, $T_d=0.10$ & 0.1500 $\pm$ 0.0022 & 0.1500 $\pm$ 0.0022 & -0.9263 $\pm$ 0.0155 & -0.5443 $\pm$ 0.0137 \\
NH, $k_{\theta}=4 \cdot 10^2$, $T_d=0.10$ & 0.1500 $\pm$ 0.0022 & 0.1500 $\pm$ 0.0022 & -0.9029 $\pm$ 0.0152 & -0.5311 $\pm$ 0.0143 \\
NH, $k_{\theta}=7 \cdot 10^2$, $T_d=0.10$ & 0.1500 $\pm$ 0.0022 & 0.1500 $\pm$ 0.0023 & -0.8947 $\pm$ 0.0151 & -0.5303 $\pm$ 0.0143 \\
NH, $k_{\theta}=10^3$, $T_d=0.10$ & 0.1500 $\pm$ 0.0022 & 0.1500 $\pm$ 0.0022 & -0.8889 $\pm$ 0.0150 & -0.5255 $\pm$ 0.0140 \\
NH, $k_{\theta}=1.4 \cdot 10^3$, $T_d=0.10$ & 0.1500 $\pm$ 0.0022 & 0.1500 $\pm$ 0.0023 & -0.8812 $\pm$ 0.0148 & -0.5195 $\pm$ 0.0140 \\
NH, $k_{\theta}=2.7 \cdot 10^3$, $T_d=0.10$ & 0.1500 $\pm$ 0.0022 & 0.1500 $\pm$ 0.0022 & -0.8506 $\pm$ 0.0150 & -0.4937 $\pm$ 0.0145 \\
NH, $k_{\theta}=5.2 \cdot 10^3$, $T_d=0.10$ & 0.1500 $\pm$ 0.0023 & 0.1500 $\pm$ 0.0023 & -0.8600 $\pm$ 0.0148 & -0.5097 $\pm$ 0.0153 \\
NH, $k_{\theta}=10^4$, $T_d=0.10$ & 0.1501 $\pm$ 0.0023 & 0.1500 $\pm$ 0.0023 & -1.0085 $\pm$ 0.0168 & -0.5949 $\pm$ 0.0191 \\
\hline
LG, $k=10 \cdot 10^3$, $T_d = 1.00$ & 0.1509 $\pm$ 0.0023 & 0.1509 $\pm$ 0.0023 & -0.9430 $\pm$ 0.0163 & -0.5889 $\pm$ 0.0136 \\
LG, $k=10 \cdot 10^3$, $T_d = 0.10$ & 0.1735 $\pm$ 0.0033 & 0.1709 $\pm$ 0.0033 & -0.8233 $\pm$ 0.0194 & -0.5204 $\pm$ 0.0148 \\
\end{tabular}
}
  \caption{Average kinetic temperature and pair energy per particle in LAMMPS simulations at $\rho=0.50$. Groups (i), (ii) and (iii) as in main text. Datasets with varying $k_{\theta}$ are generated using $k_{r}=10^4$, datasets with a unique value of $k$ are generated using $k_{\theta}=k_{r}$.}
  \label{tab:TU_0.5}
\end{table*}


\begin{table*}[h]
\resizebox{\textwidth}{!}{
  \centering
  \begin{tabular}{c|cc|cc}
    & \multicolumn{2}{c|}{$T$} & \multicolumn{2}{c}{$U$} \\
    \hline
                                & os & exp & os & exp \\
    \hline
NH, RG, $T_d=0.10$ & 0.1500 $\pm$ 0.0030 & 0.1500 $\pm$ 0.0030 & -1.2031 $\pm$ 0.0134 & -0.8423 $\pm$ 0.0127 \\
LG, RG, $T_d=1.00$ & 0.1502 $\pm$ 0.0030 & 0.1501 $\pm$ 0.0030 & -1.2023 $\pm$ 0.0139 & -0.8418 $\pm$ 0.0128 \\
LG, RG, $T_d=0.10$ & 0.1506 $\pm$ 0.0040 & 0.1505 $\pm$ 0.0038 & -1.1987 $\pm$ 0.0206 & -0.8398 $\pm$ 0.0174 \\
\hline
    \multicolumn{5}{c}{(i)} \\
\hline
NH, $k=1 \cdot 10^3$, $T_d=0.10$ & 0.1500 $\pm$ 0.0023 & 0.1500 $\pm$ 0.0022 & -1.2804 $\pm$ 0.0143 & -1.0361 $\pm$ 0.0160 \\
NH, $k=2 \cdot 10^3$, $T_d=0.10$ & 0.1500 $\pm$ 0.0023 & 0.1500 $\pm$ 0.0022 & -1.2223 $\pm$ 0.0139 & -0.9038 $\pm$ 0.0139 \\
NH, $k=4 \cdot 10^3$, $T_d=0.10$ & 0.1501 $\pm$ 0.0022 & 0.1500 $\pm$ 0.0022 & -1.1767 $\pm$ 0.0133 & -0.8088 $\pm$ 0.0131 \\
NH, $k=6 \cdot 10^3$, $T_d=0.10$ & 0.1500 $\pm$ 0.0023 & 0.1500 $\pm$ 0.0023 & -1.3081 $\pm$ 0.0144 & -0.9616 $\pm$ 0.0140 \\
NH, $k=8 \cdot 10^3$, $T_d=0.10$ & 0.1502 $\pm$ 0.0023 & 0.1501 $\pm$ 0.0023 & -1.3179 $\pm$ 0.0139 & -0.9062 $\pm$ 0.0146 \\
NH, $k=10 \cdot 10^3$, $T_d=0.10$ & 0.1501 $\pm$ 0.0023 & 0.1501 $\pm$ 0.0023 & -1.2401 $\pm$ 0.0140 & -0.8228 $\pm$ 0.0162 \\
\hline
    \multicolumn{5}{c}{(ii)} \\
\hline
NH, $k=1 \cdot 10^3$, $T_d=0.10$ & 0.1500 $\pm$ 0.0023 & 0.1500 $\pm$ 0.0022 & -1.2804 $\pm$ 0.0143 & -1.0361 $\pm$ 0.0160 \\
NH, $k=2 \cdot 10^3$, $T_d=0.05$ & 0.1500 $\pm$ 0.0022 & 0.1500 $\pm$ 0.0022 & -1.2413 $\pm$ 0.0138 & -0.9266 $\pm$ 0.0142 \\
NH, $k=4 \cdot 10^3$, $T_d=0.05$ & 0.1500 $\pm$ 0.0023 & 0.1500 $\pm$ 0.0022 & -1.2132 $\pm$ 0.0135 & -0.8642 $\pm$ 0.0132 \\
NH, $k=6 \cdot 10^3$, $T_d=0.05$ & 0.1500 $\pm$ 0.0023 & 0.1501 $\pm$ 0.0023 & -1.2942 $\pm$ 0.0141 & -0.9130 $\pm$ 0.0136 \\
NH, $k=8 \cdot 10^3$, $T_d=0.05$ & 0.1500 $\pm$ 0.0023 & 0.1500 $\pm$ 0.0023 & -1.3478 $\pm$ 0.0145 & -0.9326 $\pm$ 0.0152 \\
NH, $k=10 \cdot 10^3$, $T_d=0.01$ & 0.1500 $\pm$ 0.0022 & 0.1500 $\pm$ 0.0022 & -1.2101 $\pm$ 0.0214 & -0.8401 $\pm$ 0.0153 \\
\hline
    \multicolumn{5}{c}{(iii)} \\
\hline
NH, $k_{\theta}= 10^2$, $T_d=0.10$ & 0.1500 $\pm$ 0.0022 & 0.1500 $\pm$ 0.0023 & -1.1875 $\pm$ 0.0132 & -0.8301 $\pm$ 0.0127 \\
NH, $k_{\theta}=2 \cdot 10^2$, $T_d=0.10$ & 0.1500 $\pm$ 0.0023 & 0.1500 $\pm$ 0.0022 & -1.1979 $\pm$ 0.0135 & -0.8160 $\pm$ 0.0133 \\
NH, $k_{\theta}=4 \cdot 10^2$, $T_d=0.10$ & 0.1500 $\pm$ 0.0022 & 0.1500 $\pm$ 0.0022 & -1.1792 $\pm$ 0.0133 & -0.7967 $\pm$ 0.0141 \\
NH, $k_{\theta}=7 \cdot 10^2$, $T_d=0.10$ & 0.1500 $\pm$ 0.0023 & 0.1500 $\pm$ 0.0022 & -1.1731 $\pm$ 0.0131 & -0.7955 $\pm$ 0.0139 \\
NH, $k_{\theta}=10^3$, $T_d=0.10$ & 0.1500 $\pm$ 0.0022 & 0.1500 $\pm$ 0.0022 & -1.1682 $\pm$ 0.0135 & -0.7918 $\pm$ 0.0138 \\
NH, $k_{\theta}=1.4 \cdot 10^3$, $T_d=0.10$ & 0.1500 $\pm$ 0.0022 & 0.1500 $\pm$ 0.0022 & -1.1625 $\pm$ 0.0134 & -0.7860 $\pm$ 0.0138 \\
NH, $k_{\theta}=2.7 \cdot 10^3$, $T_d=0.10$ & 0.1500 $\pm$ 0.0023 & 0.1500 $\pm$ 0.0022 & -1.1385 $\pm$ 0.0136 & -0.7598 $\pm$ 0.0142 \\
NH, $k_{\theta}=5.2 \cdot 10^3$, $T_d=0.10$ & 0.1500 $\pm$ 0.0022 & 0.1500 $\pm$ 0.0022 & -1.1344 $\pm$ 0.0134 & -0.7594 $\pm$ 0.0144 \\
NH, $k_{\theta}= 10^4$, $T_d=0.10$ & 0.1501 $\pm$ 0.0023 & 0.1501 $\pm$ 0.0023 & -1.2399 $\pm$ 0.0138 & -0.8234 $\pm$ 0.0163 \\
\hline
LG, $k=10 \cdot 10^3$, $T_d = 1.00$ & 0.1509 $\pm$ 0.0023 & 0.1509 $\pm$ 0.0023 & -1.2103 $\pm$ 0.0141 & -0.8578 $\pm$ 0.0134 \\
LG, $k=10 \cdot 10^3$, $T_d = 0.10$ & 0.1761 $\pm$ 0.0035 & 0.1739 $\pm$ 0.0033 & -1.0878 $\pm$ 0.0183 & -0.7661 $\pm$ 0.0155 \\
\end{tabular}
}
  \caption{Average kinetic temperature and pair energy per particle in LAMMPS simulations at $ \rho=0.75$. Groups (i), (ii) and (iii) as in main text. Datasets with varying $k_{\theta}$
  are generated using $k_{r}=10^4$, datasets with a unique value of $k$ are generated using $k_{\theta}=k_{r}$.}
  \label{tab:TU_0.75}
\end{table*}

\clearpage
\section{Computational performance}
We briefly discuss here the computational performances of the different sets of parameters, not showed in the main text. We highlight the fact that, here, the disadvantage of fixing a single value for both spring constants (radial and angular) and decreasing the time step becomes evident. Indeed, the small increase in performance is negated by the fact that 5-10 times more time steps are needed (in the cases considered) to simulate comparable trajectories.

\begin{table*}[h]
  \centering
  \begin{tabular}{c|cc}
    & \multicolumn{2}{c}{ksteps per second $(s^{-1})$} \\
    \hline
        & os & exp \\
    \hline
\hline
NH, RG, $T_d=0.10$ & 2.68 $\pm$ 0.21 & 1.20 $\pm$ 0.04 \\
LG, RG, $T_d=1.00$ & 2.62 $\pm$ 0.20 & 1.19 $\pm$ 0.05 \\
LG, RG, $T_d=0.10$ & 2.62 $\pm$ 0.20 & 1.16 $\pm$ 0.03 \\
\hline
    \multicolumn{3}{c}{(i)} \\
\hline
NH, $k=1 \cdot 10^3$, $T_d=0.10$ & 5.52 $\pm$ 0.72 & 1.59 $\pm$ 0.06 \\
NH, $k=2 \cdot 10^3$, $T_d=0.10$ & 5.59 $\pm$ 0.69 & 1.61 $\pm$ 0.05 \\
NH, $k=4 \cdot 10^3$, $T_d=0.10$ & 5.64 $\pm$ 0.59 & 1.63 $\pm$ 0.03 \\
NH, $k=6 \cdot 10^3$, $T_d=0.10$ & 5.13 $\pm$ 1.32 & 1.54 $\pm$ 0.14 \\
NH, $k=8 \cdot 10^3$, $T_d=0.10$ & 5.32 $\pm$ 1.03 & 1.58 $\pm$ 0.09 \\
NH, $k=10 \cdot 10^3$, $T_d=0.10$ & 5.53 $\pm$ 0.74 & 1.62 $\pm$ 0.05 \\
\hline
    \multicolumn{3}{c}{(ii)} \\
\hline
NH, $k=1 \cdot 10^3$, $T_d=0.10$ & 5.52 $\pm$ 0.72 & 1.59 $\pm$ 0.06 \\
NH, $k=2 \cdot 10^3$, $T_d=0.05$ & 6.17 $\pm$ 1.01 & 1.75 $\pm$ 0.13 \\
NH, $k=4 \cdot 10^3$, $T_d=0.05$ & 6.01 $\pm$ 0.82 & 1.67 $\pm$ 0.04 \\
NH, $k=6 \cdot 10^3$, $T_d=0.01$ & 6.75 $\pm$ 1.31 &  \\
NH, $k=6 \cdot 10^3$, $T_d=0.05$ & & 1.64 $\pm$ 0.09 \\
NH, $k=8 \cdot 10^3$, $T_d=0.01$ & 6.57 $\pm$ 0.96 &  \\
NH, $k=8 \cdot 10^3$, $T_d=0.05$ & & 1.62 $\pm$ 0.11 \\
NH, $k=10 \cdot 10^3$, $T_d=0.01$ & 6.77 $\pm$ 1.28 & 1.93 $\pm$ 0.14 \\
\hline
\multicolumn{3}{c}{(iii)} \\
\hline
NH, $k_{\theta}=1 \cdot 10^2$, $T_d=0.10$ & 5.78 $\pm$ 0.79 & 1.67 $\pm$ 0.09 \\
NH, $k_{\theta}=2 \cdot 10^2$, $T_d=0.10$ & 5.62 $\pm$ 0.70 & 1.63 $\pm$ 0.04 \\
NH, $k_{\theta}=4 \cdot 10^2$, $T_d=0.10$ & 5.60 $\pm$ 0.66 & 1.62 $\pm$ 0.04 \\
NH, $k_{\theta}=7 \cdot 10^2$, $T_d=0.10$ & 5.64 $\pm$ 0.67 & 1.62 $\pm$ 0.04 \\
NH, $k_{\theta}=10^3$, $T_d=0.10$ & 5.63 $\pm$ 0.65 & 1.62 $\pm$ 0.04 \\
NH, $k_{\theta}=1.4 \cdot 10^3$, $T_d=0.10$ & 5.66 $\pm$ 0.64 & 1.62 $\pm$ 0.04 \\
NH, $k_{\theta}=2.7 \cdot 10^3$, $T_d=0.10$ & 5.69 $\pm$ 0.59 & 1.64 $\pm$ 0.03 \\
NH, $k_{\theta}=5.2 \cdot 10^3$, $T_d=0.10$ & 5.67 $\pm$ 0.60 & 1.62 $\pm$ 0.04 \\
NH, $k_{\theta}=10^4$, $T_d=0.10$ & 5.72 $\pm$ 0.87 & 1.67 $\pm$ 0.12 \\
\hline
LG, $k=10 \cdot 10^3$, $T_d = 1.00$ & 4.77 $\pm$ 0.58 & 1.57 $\pm$ 0.05 \\
LG, $k=10 \cdot 10^3$, $T_d = 0.10$ & 4.58 $\pm$ 0.40 & 1.55 $\pm$ 0.02 \\
\hline
\end{tabular}
  \caption{Average computational performance, measured in kilo-steps (ksteps) per second, in LAMMPS
  simulations for different models (os and exp) and $\rho=0.25$. Groups (i), (ii) and (iii) as in main text. Datasets with varying $k_{\theta}$
  are generated using $k_{r}=10^4$, datasets with a unique value of $k$ are generated using $k_{\theta}=k_{r}$.}
  \label{tab:times1}
\end{table*}

\begin{table*}[h]
  \centering
  \begin{tabular}{c|cc}
    & \multicolumn{2}{c}{ksteps per second $(s^{-1})$} \\
    \hline
        & os & exp \\
    \hline
NH, RG, $T_d=0.10$ & 2.09 $\pm$ 0.09 & 0.70 $\pm$ 0.01 \\
LG, RG, $T_d=1.00$ & 2.06 $\pm$ 0.08 & 0.70 $\pm$ 0.01 \\
LG, RG, $T_d=0.10$ & 2.06 $\pm$ 0.09 & 0.70 $\pm$ 0.01 \\
\hline
    \multicolumn{3}{c}{(i)} \\
\hline
NH, $k=1 \cdot 10^3$, $T_d=0.10$ & 3.97 $\pm$ 0.29 & 0.89 $\pm$ 0.02 \\
NH, $k=2 \cdot 10^3$, $T_d=0.10$ & 4.04 $\pm$ 0.29 & 0.89 $\pm$ 0.02 \\
NH, $k=4 \cdot 10^3$, $T_d=0.10$ & 4.06 $\pm$ 0.25 & 0.90 $\pm$ 0.03 \\
NH, $k=6 \cdot 10^3$, $T_d=0.10$ & 3.89 $\pm$ 0.41 & 0.88 $\pm$ 0.01 \\
NH, $k=8 \cdot 10^3$, $T_d=0.10$ & 4.10 $\pm$ 0.48 & 0.92 $\pm$ 0.05 \\
NH, $k=10 \cdot 10^3$, $T_d=0.10$ & 4.04 $\pm$ 0.31 & 0.90 $\pm$ 0.03 \\

\hline
    \multicolumn{3}{c}{(ii)} \\
\hline
NH, $k=1 \cdot 10^3$, $T_d=0.10$ & 3.97 $\pm$ 0.29 & 0.89 $\pm$ 0.02 \\
NH, $k=2 \cdot 10^3$, $T_d=0.05$ & 4.31 $\pm$ 0.37 & 0.92 $\pm$ 0.02 \\
NH, $k=4 \cdot 10^3$, $T_d=0.05$ & 4.33 $\pm$ 0.34 & 0.92 $\pm$ 0.02 \\
NH, $k=6 \cdot 10^3$, $T_d=0.05$ & 4.24 $\pm$ 0.40 & 0.91 $\pm$ 0.02 \\
NH, $k=8 \cdot 10^3$, $T_d=0.05$ & 4.11 $\pm$ 0.54 & 0.91 $\pm$ 0.02 \\
NH, $k=10 \cdot 10^3$, $T_d=0.01$ & 4.96 $\pm$ 0.70 & 0.99 $\pm$ 0.04 \\
\hline
    \multicolumn{3}{c}{(iii)} \\
\hline
NH, $k_{\theta}=10^2$, $T_d=0.10$ & 4.08 $\pm$ 0.28 & 0.90 $\pm$ 0.02 \\
NH, $k_{\theta}=2 10^2$, $T_d=0.10$ & 4.05 $\pm$ 0.30 & 0.90 $\pm$ 0.03 \\
NH, $k_{\theta}=4 10^2$, $T_d=0.10$ & 4.06 $\pm$ 0.27 & 0.89 $\pm$ 0.03 \\
NH, $k_{\theta}=7 \cdot 10^2$, $T_d=0.10$ & 4.05 $\pm$ 0.27 & 0.89 $\pm$ 0.03 \\
NH, $k_{\theta}= 10^3$, $T_d=0.10$ & 4.07 $\pm$ 0.27 & 0.89 $\pm$ 0.03 \\
NH, $k_{\theta}=1.4 \cdot 10^3$, $T_d=0.10$ & 4.07 $\pm$ 0.27 & 0.89 $\pm$ 0.03 \\
NH, $k_{\theta}=2.7 \cdot 10^3$, $T_d=0.10$ & 4.07 $\pm$ 0.25 & 0.89 $\pm$ 0.03 \\
NH, $k_{\theta}=5,2 \cdot 10^3$, $T_d=0.10$ & 4.08 $\pm$ 0.25 & 0.89 $\pm$ 0.03 \\
NH, $k_{\theta}= 10^4$, $T_d=0.10$ & 4.03 $\pm$ 0.30 & 0.90 $\pm$ 0.02 \\
\hline
LG, $k=10 \cdot 10^3$, $T_d = 1.00$ & 3.67 $\pm$ 0.24 & 0.88 $\pm$ 0.02 \\
LG, $k=10 \cdot 10^3$, $T_d = 0.10$ & 3.49 $\pm$ 0.15 & 0.87 $\pm$ 0.03 \\
\hline
\end{tabular}
  \caption{Average computational performance, measured in kilo-steps (ksteps) per second, in LAMMPS
  simulations for different models (os and exp) and $\rho=0.50$. Groups (i), (ii) and (iii) as in main text. Datasets with varying $k_{\theta}$
  are generated using $k_{r}=10^4$, datasets with a unique value of $k$ are generated using $k_{\theta}=k_{r}$.}
  \label{tab:times2}
\end{table*}

\begin{table*}[h]
  \centering
  \begin{tabular}{c|cc}
    & \multicolumn{2}{c}{ksteps per second $(s^{-1})$} \\
    \hline
        & os & exp \\
    \hline
NH, RG, $T_d=0.10$ & 1.69 $\pm$ 0.02 & 0.50 $\pm$ 0.02 \\
LG, RG, $T_d=1.00$ & 1.68 $\pm$ 0.02 & 0.49 $\pm$ 0.02 \\
LG, RG, $T_d=0.10$ & 1.68 $\pm$ 0.02 & 0.49 $\pm$ 0.03 \\
\hline
    \multicolumn{3}{c}{(i)} \\
\hline
NH, $k=1 \cdot 10^3$, $T_d=0.10$ & 3.05 $\pm$ 0.06 & 0.62 $\pm$ 0.03 \\
NH, $k=2 \cdot 10^3$, $T_d=0.10$ & 3.05 $\pm$ 0.04 & 0.62 $\pm$ 0.03 \\
NH, $k=4 \cdot 10^3$, $T_d=0.10$ & 3.07 $\pm$ 0.05 & 0.62 $\pm$ 0.03 \\
NH, $k=6 \cdot 10^3$, $T_d=0.10$ & 3.05 $\pm$ 0.05 & 0.62 $\pm$ 0.03 \\
NH, $k=8 \cdot 10^3$, $T_d=0.10$ & 3.07 $\pm$ 0.07 & 0.62 $\pm$ 0.03 \\
NH, $k=10 \cdot 10^3$, $T_d=0.10$ & 3.06 $\pm$ 0.06 & 0.62 $\pm$ 0.03 \\
\hline
    \multicolumn{3}{c}{(ii)} \\
\hline
NH, $k=1 \cdot 10^3$, $T_d=0.10$ & 3.05 $\pm$ 0.06 & 0.62 $\pm$ 0.03 \\
NH, $k=2 \cdot 10^3$, $T_d=0.05$ & 3.26 $\pm$ 0.07 & 0.63 $\pm$ 0.03 \\
NH, $k=4 \cdot 10^3$, $T_d=0.05$ & 3.28 $\pm$ 0.05 & 0.63 $\pm$ 0.03 \\
NH, $k=6 \cdot 10^3$, $T_d=0.05$ & 3.28 $\pm$ 0.04 & 0.63 $\pm$ 0.03 \\
NH, $k=8 \cdot 10^3$, $T_d=0.05$ & 3.26 $\pm$ 0.05 & 0.63 $\pm$ 0.03 \\
NH, $k=10 \cdot 10^3$, $T_d=0.01$ & 3.64 $\pm$ 0.14 & 0.66 $\pm$ 0.06 \\
\hline
    \multicolumn{3}{c}{(iii)} \\
\hline
NH, $k_{\theta}=10^2$, $T_d=0.10$ & 3.05 $\pm$ 0.06 & 0.62 $\pm$ 0.03 \\
NH, $k_{\theta}=2 \cdot 10^2$, $T_d=0.10$ & 3.06 $\pm$ 0.04 & 0.62 $\pm$ 0.03 \\
NH, $k_{\theta}=4 \cdot 10^2$, $T_d=0.10$ & 3.07 $\pm$ 0.04 & 0.61 $\pm$ 0.03 \\
NH, $k_{\theta}=7 \cdot 10^2$, $T_d=0.10$ & 3.05 $\pm$ 0.05 & 0.62 $\pm$ 0.03 \\
NH, $k_{\theta}=10^3$, $T_d=0.10$ & 3.11 $\pm$ 0.09 & 0.63 $\pm$ 0.05 \\
NH, $k_{\theta}=1.4 \cdot 10^3$, $T_d=0.10$ & 3.07 $\pm$ 0.04 & 0.61 $\pm$ 0.03 \\
NH, $k_{\theta}=2.7 \cdot 10^3$, $T_d=0.10$ & 3.05 $\pm$ 0.05 & 0.62 $\pm$ 0.03 \\
NH, $k_{\theta}=5.2 \cdot 10^3$, $T_d=0.10$ & 3.07 $\pm$ 0.05 & 0.61 $\pm$ 0.03 \\
NH, $k_{\theta}= 10^4$, $T_d=0.10$ & 3.07 $\pm$ 0.06 & 0.62 $\pm$ 0.04 \\
\hline
LG, $k=10 \cdot 10^3$, $T_d = 1.00$ & 2.92 $\pm$ 0.06 & 0.62 $\pm$ 0.04 \\
LG, $k=10 \cdot 10^3$, $T_d = 0.10$ & 2.80 $\pm$ 0.07 & 0.62 $\pm$ 0.05 \\
\hline
\end{tabular}
  \caption{Average computational performance, measured in kilo-steps (ksteps) per second, in LAMMPS
  simulations for different models (os and exp) and $\rho=0.75$. Groups (i), (ii) and (iii) as in main text. Datasets with varying $k_{\theta}$
  are generated using $k_{r}=10^4$, datasets with a unique value of $k$ are generated using $k_{\theta}=k_{r}$.}
  \label{tab:times3}
\end{table*}

\end{document}